\documentclass[lettersize,journal]{IEEEtran}
\usepackage[utf8]{inputenc}
\usepackage[colorlinks=true,
            linkcolor=black,
            urlcolor=black,
            citecolor=black]{hyperref}
\usepackage{amsmath,amsfonts}
\usepackage{algorithm}
\usepackage{algpseudocode}
\usepackage{array}
\usepackage{makecell}
\usepackage{textcomp}
\usepackage[caption=false,font=footnotesize]{subfig}
\usepackage{booktabs}
\usepackage{threeparttable}
\usepackage{orcidlink}
\usepackage{tabularx}
\usepackage{stfloats}
\usepackage{placeins}
\usepackage{multirow}
\usepackage{soul}
\usepackage{xcolor}
\setulcolor {red}
\usepackage{arydshln}
\usepackage{url}
\usepackage{verbatim}
\usepackage{graphicx}   
\usepackage{adjustbox}    
\usepackage{cite}
\newcolumntype{C}[1]{>{\centering\arraybackslash}p{#1}}
\begin{document}

\title{GMA-SAWGAN-GP: A Novel Data Generative Framework to Enhance IDS Detection Performance}

\author{%
  Ziyu~Mu\orcidlink{0009-0008-0697-7580}, %
  Xiyu~Shi\orcidlink{0000-0001-6174-3383},~\IEEEmembership{Senior Member,~IEEE}, %
  Safak~Dogan\orcidlink{0000-0002-1465-6495},~\IEEEmembership{Senior Member,~IEEE}%
  \thanks{All authors are with the  Institute for Digital Technologies, Loughborough University London, London E20 3BS, UK.}%
  \thanks{Corresponding author: Ziyu Mu (e-mail: z.mu@lboro.ac.uk).}%
}



\maketitle

\begin{abstract}
Intrusion Detection System (IDS) is often calibrated to known attacks and generalizes poorly to unknown threats. This paper proposes GMA-SAWGAN-GP, a novel generative augmentation framework built on a Self-Attention-enhanced Wasserstein GAN with Gradient Penalty (WGAN-GP). The generator employs Gumbel–Softmax regularization to model discrete fields, while a Multilayer Perceptron (MLP)-based AutoEncoder acts as a manifold regularizer. A lightweight gating network adaptively balances adversarial and reconstruction losses via entropy regularization, improving stability and mitigating mode collapse. The self-attention mechanism enables the generator to capture both short- and long-range dependencies among features within each record while preserving categorical semantics through Gumbel–Softmax heads. Extensive experiments on NSL-KDD, UNSW-NB15, and CICIDS2017 using five representative IDS models demonstrate that GMA-SAWGAN-GP significantly improves detection performance on known attacks and enhances generalization to unknown attacks. Leave-One-Attack-type-Out (LOAO) evaluations using Area Under the Receiver Operating Characteristic (AUROC) and True Positive Rate at a 5\% False Positive Rate (TPR@5\%FPR) confirm that IDS models trained on augmented datasets achieve higher robustness under unseen attack scenarios. Ablation studies validate the contribution of each component to performance gains. Compared with baseline models, the proposed framework improves binary classification accuracy by an average of 5.3\% and multi-classification accuracy by 2.2\%, while AUROC and TPR@5\%FPR for unknown attacks increase by 3.9\% and 4.8\%, respectively, across the three datasets. Overall, GMA-SAWGAN-GP provides an effective approach to generative augmentation for mixed-type network traffic, improving IDS accuracy and resilience.
\end{abstract}


\section{Introduction}
\IEEEPARstart{A}{n} Intrusion Detection System (IDS) is a critical defensive layer in modern networks, continuously monitoring network traffic and endpoint behavior to identify known threats, suspicious activities, and policy violations. In practice, security teams face two persistent challenges: weak calibration to the distribution of known attacks and limited generalization to unknown attacks \cite{alshammari2025smart}. Expanding labeled datasets or updating signature-based knowledge bases offer incremental improvements, but these approaches are costly, slow to deploy, and degrade as network traffic evolves\cite{tao2025accuracy}.

Data generative augmentation based on Generative Adversarial Networks (GANs) \cite{goodfellow2014generativeadversarialnetworks} offers a promising alternative. By synthesizing training flows, machine learning-based IDS models can explore a richer sample space and strengthen representation and discrimination. However, existing augmentation pipelines for network intrusion detection face three major limitations. First, feature representation is mixed: fields such as protocol and service are discrete, while several others are continuous. Prior work has either utilized one-hot encoding \cite{9512057} on discrete fields or removed them entirely \cite{suresh2024deep, luqman2025intelligent, xu2025multimodal}. Both approaches are problematic. One-hot encoding causes severe dimensional growth and sparsity in high-cardinality fields, erases semantic proximity among categories, struggles with long-tail and unseen values, and its non-differentiable nature complicates end-to-end generative training. Dropping discrete features discards valuable signals, weakening detection and calibration \cite{liang2025efficient, mumtaz2022hierarchy, majidian2024optimizing}. Second, feature dependencies are critical. Real network flows exhibit short- and long-range relationships among features within a record that basic multilayer perceptrons fail to capture. Third, adversarial generative models are prone to mode collapse during training, leading to instability, overfitting, and inadequate coverage of minority patterns, which undermines detection improvements \cite{wang2024application, jabbar2021survey}.

This paper proposes a novel generative augmentation model, GMA-SAWGAN-GP, designed to enhance the detection of known attacks while improving robustness against unknown attacks in IDS models. GMA-SAWGAN-GP is designed to model both discrete and continuous features in network traffic through three key components. First, discrete-aware synthesis is achieved in the Generator (G) via Gumbel-Softmax regularizer\cite{shah2024improving} applied to per-feature codebooks for categorical fields, alongside a tanh head for continuous features. Second, G is trained under the Wasserstein GAN with Gradient Penalty (WGAN-GP) framework \cite{gulrajani2017improvedtrainingwassersteingans} and equipped with a feature-wise Self-Attention (SA) mechanism, enabling attention among features and capturing both short- and long-range dependencies in network traffic. Third, an AutoEncoder (AE) \cite{bank2021autoencoders} serves as a manifold regularizer, pulling synthetic samples toward the real data manifold and stabilizes training. Additionally, an entropy-based loss attention gate \cite{mao2023crossentropylossfunctionstheoretical} adaptively balances reconstruction and adversarial objectives during training, improving optimization stability, broadening coverage of rare but security-critical patterns, and reducing mode collapse without manual reweighting.

In this work, evaluation is conducted on three publicly available network traffic datasets, namely NSL-KDD\cite{5356528}, UNSW-NB15 \cite{7348942}, and CICIDS2017\cite{sharafaldin2018toward}, using five representative IDS models. For known attacks, detection performance is reported as the mean and standard deviation of accuracy and macro F1-score across multiple runs. Generalization to unknown attacks is assessed with the Leave-One-Attack-type-Out (LOAO) method, with Area Under the Receiver Operating Characteristic (AUROC) and True Positive Rate at a 5\% False Positive Rate (TPR@5\%FPR) as primary metrics. During the training of G, the Sliced Wasserstein Distance (SWD) is monitored to quantify distributional fidelity.

To summarize, the key contributions of this work are:
\begin{itemize}

\item A discrete-aware generative augmentation method is developed, which preserves categorical semantics through Gumbel-Softmax regularization while remaining aligned with continuous features. Learnable codebooks with temperature annealing are employed to stabilize discrete modeling and reconstruction.

\item A feature-wise self-attention mechanism is integrated into the generator within the WGAN-GP, explicitly modeling cross-feature dependencies and improving realism without hand-crafting feature groups.

\item An autoencoder-guided manifold regularization is incorporated by enforcing reconstruction consistency on generated flows through a frozen Encoder and Decoder during the generator update, which constrains synthesis toward reconstructible local geometry and alleviates mode collapse.

\item An entropy-based adaptive loss attention gating network is devised, which dynamically balances reconstruction and adversarial objectives, yielding stable training and broader coverage of minority patterns critical for intrusion detection.

\end{itemize}

The remainder of the paper is organized as follows. Section II reviews current intrusion detection methods and prior GAN-based approaches. Section III presents the proposed model architecture in detail. Section IV describes the experimental setup, including the datasets, the five representative IDS models, and the sample generation procedure. Section V presents the evaluation results for binary classification, multi-classification, and LOAO experiments, compares them with the state-of-the-art GAN-based data generative models, and discusses the ablation study and limitations. Section VI concludes the paper and outlines directions for future work.

\section{Related Work}

A large number of studies have used machine learning and deep learning to build IDS models. Zou et al. combined a Decision Tree (DT) strategy with Support Vector Machines (SVM) and reported multi-classification accuracies of 85.95\% on the NSL-KDD dataset and 81.21\% on the UNSW-NB15 dataset\cite{10057402}. Momand et al. \cite{momand2024abcnn} integrated an attention mechanism with a Convolutional Neural Network (CNN) and reached 99.52\% accuracy on the CICIDS2017 dataset for multi-classification. A Likelihood Naive Bayes classifier with a pigeon-inspired metaheuristic was proposed in \cite{9765958}, reporting binary classification accuracies of 99.45\% on CICIDS2017 and 99.99\% NSL-KDD. Hassouneh \& Al-Sharaeh evaluated Long Short-Term Memory (LSTM) variants as the IDS models for binary detection on UNSW-NB15 dataset, with a Bidirectional LSTM (BiLSTM)-based IDS model reaching 95.01\% accuracy with Synthetic Minority Oversampling Technique (SMOTE) and 94.68\% without SMOTE\cite{hassouneh2025intrusion}. Kumar et al. \cite{10968274} compared classical machine learning-based IDS models on UNSW-NB15 and found that, for multi-classification, a Random Forest-based IDS achieved 98.63\% accuracy, outperforming Logistic Regression and DT models. Karthikeyan et al. proposed a Firefly algorithm-based feature selection method combined with Grey Wolf Optimization-tuned SVM, evaluated on multi-classification detection using the NSL-KDD dataset, and reported an average training accuracy of 99.34\% \cite{karthikeyan2024firefly}. These methods perform well on known, labeled distributions but typically optimize for aggregate accuracy and rely on static training-testing splits, providing limited insight into robustness against distribution shift, rare minority classes, or novel attacks.

Building on this, recent studies have employed generative models - particularly GAN-based approaches -to synthesize network traffic and subsequently train IDS models on augmented datasets. Rahman et al. introduced SYN-GAN \cite{rahman2024syn} and demonstrated that IDS models based on Gaussian Naive Bayes and K-Nearest Neighbors, when trained exclusively on fully synthetic data, achieved binary accuracies of 90\% and 84\% on UNSW-NB15 and NSL-KDD, respectively. Li et al. proposed Denoise Autoencoder GAN (DAE-GAN) \cite{9893038}, which includes multiple denoising autoencoders with a binary Multilayer Perceptron (MLP) discriminator as the IDS model, reporting 98.6\% precision on NSL-KDD and 98.5\% on UNSW-NB15 for binary classification. Arafah et al. presented AE-WGAN \cite{arafah2025anomaly}, combining a Gated Recurrent Unit and CNN-based IDS model, achieving binary accuracies of 98\% and 93\% on NSL-KDD and CICIDS2017, respectively. They further introduced E-BiGAN \cite{arafah2025enhanced}by pairing an AE model with a Bidirectional GAN, reaching 98\% binary accuracy with an LSTM-based IDS on both NSL-KDD and CICIDS2017. Wang and Kandah proposed a multi-critic WGAN-GP \cite{10852156}, specifically targeting class imbalance by generating 10,000 minority-class samples for R2L and U2R attacks for the NSL-KDD dataset, and trained a DNN-based IDS to achieve 81\% multi-class accuracy.  Ding et al. introduced Tabular Multi-Generator GAN (TMG-GAN) \cite{10312801}, a multi-generator architecture with a cosine-similarity objective, reporting near-perfect macro precision, recall, and F1 scores ($\approx 99.7\%$) for binary classification and strong multi-class performance on CICIDS2017, though results on UNSW-NB15 were notably lower. Finally, Srivastava et al. combined a Wasserstein Conditional Generative Adversarial Network with Gradient Penalty (WCGAN-GP) and genetic algorithm-based feature selection \cite{srivastava2023wcgan}, leveraging an XGBoost-based IDS  \cite{Chen_2016} to achieve competitive multi-class accuracies of 95.54\% and 89.58\% on NSL-KDD and UNSW-NB15, respectively. While promising, most of these studies either treat discrete fields with one-hot encoding or ignore them altogether, emphasize accuracy over metrics such as F1-score and false positive rate, and rarely evaluate generalization to previously unseen attacks using the LOAO method. Consequently, the interaction between mixed discrete and continuous features and the challenge of detecting unknown attacks remains insufficiently explored.

Feature engineering remains a common practice in intrusion detection for selecting relevant features during training and testing. Zorarpacı employed a wrapper-based feature selection, retaining 7 and 13 features for the NSL-KDD training and testing sets, respectively, and 13 for UNSW-NB15 \cite{zorarpaci2024fast}. Using an ensemble classifier-based IDS model, the approach achieved 97.4\% multi-classification accuracy on the NSL-KDD testing set and 82.7\% on UNSW-NB15. Thakkar and Lohiya proposed an importance-based method \cite{thakkar2023fusion}, selecting 21 features for NSL-KDD, 21 for UNSW-NB15, and 64 for CICIDS2017; a DNN-based IDS model achieved 99.84\%, 89.03\%, and 99.80\% in multi-classification, respectively. Zeng et al. introduced Causal Grey Wolf Optimization with an enhanced convolutional autoencoder (CGWO-ECAE) model \cite{zeng2025causal}, selecting 14 features for NSL-KDD and 10 for UNSW-NB15, and used a softmax classifier-based IDS model, reporting 99.91\% and 99.82\% accuracy for binary and multi-class classification on NSL-KDD, and 96.21\% and 94.18\% on UNSW-NB15. These results underscore the value of feature engineering; however, most pipelines rely on labeled, in-distribution data and static feature subsets, and adopt categorical encodings that increase dimensionality and discard semantic relationships. Consequently, feature selection alone does not address the need for generative augmentation that preserves mixed-type semantics and supports evaluation under unknown attack scenarios.

Against this backdrop, the present work introduces a discrete-aware generator designed to address mixed discrete and continuous flows, mitigating sparsity and vanishing gradient issues caused by one-hot encoding. In addition, it integrates an explicit cross-feature dependency module to capture interactions between discrete and continuous features. Finally, the proposed approach evaluates generalization to unseen attacks using the LOAO method, thereby addressing critical gaps in prior IDS models, generative augmentation strategies, and feature selection techniques.

\section{Proposed Methodology}

\begin{figure*}[htbp]
  \centering
  \includegraphics[width=0.85\textwidth]{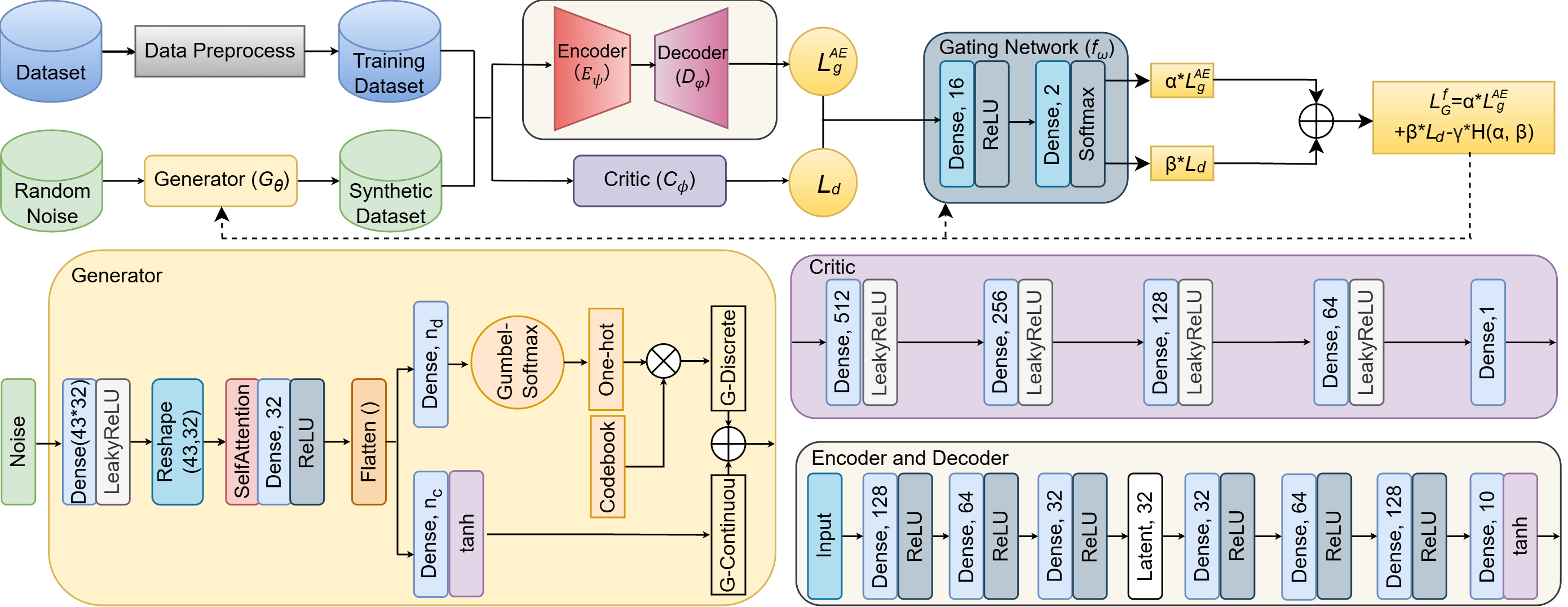}
  \caption {The architecture of GMA-SAWGAN-GP: Upper part shows the workflow of the proposed method, while the lower part illustrates the structures of the Generator, Critic, and AutoEncoder blocks. $\alpha,\beta$ are adaptive mixing weights output by the gating network $f_{\omega}$ to balance $L_{g}^{AE}$ and $L_d$.}
  \label{fig:GMA_SAWGANGP}
\end{figure*}

Fig. \ref{fig:GMA_SAWGANGP} illustrates the proposed GMA-SAWGAN-GP architecture. Within the WGAN-GP framework, the generator $G_{\theta}$ is trained on mixed discrete and continuous network flow features. Each flow record is encoded as an $F$-dimensional feature vector, where $F$ denotes the number of features in the preprocessed dataset, and $G_{\theta}$ applies a feature-wise SA mechanism to capture intra-record dependencies. Categorical attributes are generated as scalars using a straight-through Gumbel-Softmax regularized over per-feature codebooks, while continuous attributes are produced by a dense head with \(\tanh\) activation. Training alternates between five Critic ($C_{\phi}$) steps and one generator step, incorporating a Gradient Penalty (GP) on interpolations. A shared AE enforces reconstruction consistency on generated samples, and a lightweight two-layer gating mechanism adaptively balances the reconstruction term with the adversarial objective. Details of the architecture, loss functions, and pseudocode are provided in the following sections.

\subsection{WGAN-GP with Feature-wise SA mechanism}

\begin{algorithm}[t]
\caption{WGAN-GP with Feature-wise Self-Attention, Discrete-Aware Generator}
\label{alg:wgangp-col-sa}
\begin{algorithmic}[1]
\Require Data $x\!\sim\!P_{\mathrm{data}}$, prior $z\!\sim\!P_{z}$, codebooks $\{\mathbf{v}^{(c)}\}$; batch $B$, critic steps $n_c$, GP weight $\lambda$, temperature $\tau$, learning rates $\eta_C,\eta_G$
\State Init critic $C_{\phi}$, generator $G_{\theta}$ (feature-wise self-attention + discrete and continuous heads)
\For{each epoch}
  \For{each mini-batch}
    \For{$t=1$ \textbf{to} $n_c$} \Comment{Critic update (WGAN-GP)}
      \State Sample $\{x_i\}_{1}^{B}\!\sim\!P_{\mathrm{data}}$, $\{z_i\}_{1}^{B}\!\sim\!P_{z}$; set $\tilde{x}_i=G_{\theta}(z_i)$
      \State Sample $\epsilon_i\!\sim\!\mathcal{U}(0,1)$; set $\hat{x}_i=\epsilon_i x_i+(1-\epsilon_i)\tilde{x}_i$
      \State $GP=\lambda\cdot\frac{1}{B}\!\sum_{i=1}^{B}\!\big(\|\nabla_{\hat{x}_i}C_{\phi}(\hat{x}_i)\|_2-1\big)^2$
      \State $L_C=\frac{1}{B}\!\sum_{i=1}^{B}\!\big[C_{\phi}(\tilde{x}_i)-C_{\phi}(x_i)\big]+GP$;\quad $\phi\leftarrow\phi-\eta_C\nabla_{\phi} L_C$
    \EndFor
    \State Sample $\{z_i\}_{1}^{B}\!\sim\!P_{z}$ \Comment{Generator adversarial update}
   \State \textit{Feature-wise SA in $G_{\theta}$:}
        $z_i\!\xrightarrow{\mathrm{Dense}}\!\mathbf{H}_i\!\in\!\mathbb{R}^{F\times d_m}$;\;
        $\mathbf{Q}=\mathbf{H}_i\mathbf{W}_Q,\;\mathbf{K}=\mathbf{H}_i\mathbf{W}_K,\;\mathbf{V}=\mathbf{H}_i\mathbf{W}_V$
    \State $\mathrm{Attn}=\mathrm{softmax}\!\Big(\tfrac{\mathbf{Q}\mathbf{K}^{\top}}{\sqrt{d_k}}\Big)\mathbf{V}$
    \State \textit{Discrete head:} for each field $c$:
        $\mathbf{g}^{(c)}\!\sim\!\mathrm{Gumbel}(0,1)^{K_c}$,\;
        $\mathbf{y}_{\text{soft}}^{(c)}=\mathrm{softmax}\!\Big(\tfrac{\ell^{(c)}+\mathbf{g}^{(c)}}{\tau}\Big)$,\;
        $\tilde{\mathbf{y}}^{(c)}=\mathrm{one\_hot}\!\Big(\arg\max_k \mathbf{y}_{\text{soft},k}^{(c)}\Big)$,\;
        $x^{(c)}=\langle \tilde{\mathbf{y}}^{(c)},\mathbf{v}^{(c)}\rangle$
    \State \textit{Continuous head:} linear $+\tanh$ to $[-1,1]$;\; concatenate to get $\tilde{x}_i=G_{\theta}(z_i)$
    \State $L_G=-\frac{1}{B}\!\sum_{i=1}^{B}\!C_{\phi}(\tilde{x}_i)$;\quad $\theta\leftarrow\theta-\eta_G\nabla_{\theta} L_G$
    \State \textit{(Optional) record:} critic gap, $\|\nabla_{\hat{x}}C_{\phi}(\hat{x})\|_2$, $-C_{\phi}(\tilde{x})$
  \EndFor
\EndFor
\end{algorithmic}
\end{algorithm}

Weight clipping in the original WGAN \cite{arjovsky2017wassersteingan} can restrict the capacity of $C_{\phi}$ and destabilize training. WGAN-GP addresses this using the Wasserstein distance and replacing clipping with a GP that softly enforces the 1-Lipschitz constraint \cite{8167183} by driving the $C_{\phi}$’s gradient norm toward 1. This modification improves training stability and reduces mode collapse. The overall WGAN-GP training procedure with feature-wise self-attention and the discrete-aware generator is presented in Algorithm~\ref{alg:wgangp-col-sa}.

\paragraph{Basics of WGAN-GP}
Let $x\!\sim\!P_{\text{data}}$ denote a real sample, $z\!\sim\!P_{z}$ a latent vector (e.g., $z\!\sim\!\mathcal{N}(0,I)$), $\tilde{x}=G_{\theta}(z)$ the sample produced by the generator $G_{\theta}$, and $C_{\phi}(\cdot)$ the Wasserstein critic parameterized by $\phi$ that assigns a real-valued score to an input sample.
To compute the gradient penalty, we follow \cite{gulrajani2017improvedtrainingwassersteingans} and define interpolated samples $\hat{x}$ on straight lines between real and generated samples as: 
\begin{equation}
\hat{x} \;=\; \epsilon\,x + (1-\epsilon)\,\tilde{x},
\quad \epsilon \sim \mathcal{U}(0.0,1.0),
\label{eq:xhat}
\end{equation}
where $\mathcal{U}(0.0,1.0)$ denotes the continuous uniform distribution on the interval [0.0, 1.0]. Thus, $\hat{x}$ is uniformly sampled along the line segment between $x$ and $\tilde{x}$. The GP is defined as:
\begin{equation}
GP \;=\; \lambda \,\mathbb{E}_{\hat{x}}\!\left[\left(\,\big\|\nabla_{\hat{x}} C_{\phi}(\hat{x})\big\|_{2}-1\right)^{2}\right],
\label{eq:gp}
\end{equation}
where $\lambda$ is the penalty weight, $\nabla_{\hat{x}} C_{\phi}(\hat{x})$ is the gradient of $C_{\phi}$ with respect to $\hat{x}$, and $\|\cdot\|_2$ is the Euclidean norm.
The critic and generator losses are:
\begin{equation}
L_{C} = \mathbb{E}_{z\sim P_{z}}\!\big[C_{\phi}(G_{\theta}(z))\big]
      - \mathbb{E}_{x\sim P_{\text{data}}}\!\big[C_{\phi}(x)\big]
      + GP,
\label{eq:ld-critic}
\end{equation}

\begin{equation}
L_{G} = -\,\mathbb{E}_{z\sim P_{z}}\!\big[C_{\phi}(G_{\theta}(z))\big].
\label{eq:lg-gen}
\end{equation}
In the generative experiments, each generator update is preceded by five critic updates with $G_{\theta}$ fixed.

\paragraph{Adversarial Loss Function}
Each traffic record is represented as one row with $F$ features.
A parameterized $G_{\theta}$ maps $z\!\sim\!P_{z}$ to $\tilde{x}=G_{\theta}(z)$, and a parameterized critic $C_{\phi}$ assigns real-valued scores.
The adversarial loss used to train $G_{\theta}$ is defined as:
\begin{equation}
L_d \;=\; -\,\mathbb{E}_{z\sim P_{z}}\!\big[C_{\phi}(G_{\theta}(z))\big],
\label{eq:ld}
\end{equation}
which coincides with the same objective in \eqref{eq:lg-gen}.

\paragraph{Feature-wise Self-Attention Mechanism}
To model dependencies among the $F$ features within the same record, the $G_{\theta}$ applies a feature-wise SA mechanism.
Starting from $z$, a dense projection produces a hidden table $\mathbf{H}\!\in\!\mathbb{R}^{F\times d_m}$ of token embeddings, where $d_m$ is the embedding dimension.
Queries, keys, and values are calculated via learned projections $\mathbf{W}_Q,\mathbf{W}_K,\mathbf{W}_V$ as:
\begin{equation}
\mathbf{Q}=\mathbf{H}\mathbf{W}_Q,
\mathbf{K}=\mathbf{H}\mathbf{W}_K,
\mathbf{V}=\mathbf{H}\mathbf{W}_V,
\label{eq:attnQKV}
\end{equation}
and attention is obtained by:
\begin{equation}
\mathrm{Attn}(\mathbf{Q},\mathbf{K},\mathbf{V})
=\mathrm{softmax}\Big(\tfrac{\mathbf{Q}\mathbf{K}^{\top}}{\sqrt{d_k}}\Big)\mathbf{V},
\label{eq:attn}
\end{equation}
where $d_k$ is the key dimension and $(\cdot)^{\top}$ denotes transpose.

\paragraph{Discrete-aware and Continuous heads}
The outputs of $G_{\theta}$ are partitioned into a discrete head and a continuous head.
For each discrete field $c$ with $K_c$ legal values, the discrete head produces logits $\ell^{(c)}\!\in\!\mathbb{R}^{K_c}$ and draws a Gumbel-Softmax sample.
With $\tau$ being the temperature parameter and g$\sim\mathrm{Gumbel}(0,1)^{K_c}$, let:
\begin{equation}
\mathbf{y}_{\text{soft}}^{(c)} \;=\; \mathrm{softmax}\Big(\tfrac{\ell^{(c)} + \mathbf{g}}{\tau}\Big),
\label{eq:gumbel}
\end{equation}
and its one-hot encoding $\tilde{\mathbf{y}}^{(c)}=\mathrm{one\_hot}\big(\arg\max \mathbf{y}_{\text{soft}}^{(c)}\big)$, which is used in the forward pass to form a discrete output. Under the straight-through estimator \cite{jang2017categoricalreparameterizationgumbelsoftmax}, gradients of the generator objective are backpropagated through $\mathbf{y}_{\text{soft}}^{(c)}$ and the logits $\ell^{(c)}$ to update $G_{\theta}$.
This one-hot vector is mapped to a legal code $x^{(c)}$ via a fixed, field-specific codebook $\mathbf{v}^{(c)} \in \mathbb{R}^{K_c}$, obtained during preprocessing:
\begin{equation}
x^{(c)} \;=\; \langle \tilde{\mathbf{y}}^{(c)},\, \mathbf{v}^{(c)} \rangle,
\label{eq:codebook}
\end{equation}
so each categorical field contributes a single scalar to the output.
For continuous features, a linear layer followed by $\tanh$ constrains values to the normalized range $[-1.0,1.0]$.

\subsection{The AE Model and Self-Adaptive Loss}

\begin{algorithm}[t]
\caption{Reconstruction Consistency \& Self-Adaptive Balancing (AE + Gate)}
\label{alg:part2-ae-balance}
\begin{algorithmic}[1]
\Require Mini-batch $\{x_i\}_{1}^{B}\!\sim\!P_{\mathrm{data}}$, $\{z_i\}_{1}^{B}\!\sim\!P_{z}$; models $G_{\theta},C_{\phi},E_{\psi},D_{\varphi},f_{\omega}$; clip bounds $[a,b]$; entropy weight $\gamma$; learning rates $\eta_{\text{AE}},\eta_G$
\State \textbf{Forward:} $\tilde{x}_i = G_{\theta}(z_i)$;\quad $\hat{x}_i=D_{\varphi}(E_{\psi}(x_i))$;\quad $\hat{\tilde{x}}_i=D_{\varphi}(E_{\psi}(\tilde{x}_i))$
\State \textbf{AE (real-only) loss \& update:}\; $L_{\text{$r$}}^{\text{$AE$}}=\frac{1}{B}\!\sum_{i=1}^{B}\|x_i-\hat{x}_i\|_2^2$;\quad $(\psi,\varphi)\leftarrow(\psi,\varphi)-\eta_{\text{AE}}\nabla L_{r}^{AE}$
\State \textbf{AE constraint on fakes (freeze AE):}\; $L_{\text{$g$}}^{\text{$AE$}}=\frac{1}{B}\!\sum_{i=1}^{B}\|\tilde{x}_i-\hat{\tilde{x}}_i\|_2^2$
\State \textbf{Adversarial term (freeze critic):}\; $L_d=-\frac{1}{B}\!\sum_{i=1}^{B} C_{\phi}(\tilde{x}_i)$
\State \textbf{Gate:}\; $(u,v)=f_{\omega}([L_{g}^{AE},\,L_d])$;\quad $(\alpha,\beta)=\mathrm{clip}_{[a,b]}(\mathrm{softmax}([u,v]))$
\State \textbf{Final loss:}\; $L_G^{f}=\alpha L_{g}^{AE}+\beta L_d-\gamma\,H(\alpha,\beta)$
\State \textbf{Update:}\; $\theta,\omega \leftarrow \theta,\omega - \eta_G \nabla_{\{\theta,\omega\}} L_G^{f}$
\end{algorithmic}
\end{algorithm}

Relying solely on the adversarial loss $L_d$ can lead to mode collapse during WGAN-GP training. To mitigate this, an autoencoder is introduced to enforce reconstruction consistency, ensuring that synthetic samples remain close to a reconstructible local geometry. Furthermore, a learnable gating mechanism adaptively balances the reconstruction and adversarial objectives throughout training, improving stability and reducing collapse. The reconstruction consistency and self-adaptive balancing procedure are outlined in Algorithm~\ref{alg:part2-ae-balance}.

The AE consists of an encoder $E_{\psi}$ and a decoder $D_{\varphi}$. 
Reconstructions are defined as:
\begin{equation}
{x}'=D_{\varphi}\big(E_{\psi}(x)\big),
\qquad 
{\tilde{x}'}=D_{\varphi}\big(E_{\psi}(\tilde{x})\big).
\end{equation}
The AE is trained on real data using the reconstruction loss:
\begin{equation}
L_{r}^{AE}
= \mathbb{E}_{x\sim P_{\text{data}}}\!\big[\|x-{x}'\|_2^2\big].
\end{equation}
During the $G_{\theta}$ update, AE weights are frozen, and a G-side reconstruction loss is computed on generated samples:
\begin{equation}
L_{g}^{AE}
= \mathbb{E}_{z\sim P_{z}}\!\big[\|\tilde{x}-\tilde{x}'\|_2^2\big].
\end{equation}

The adversarial loss in (\ref{eq:ld}) aligns the objective of $G_{\theta}$ in (\ref{eq:lg-gen}) for standard WGAN-GP. 
To combine the adversarial and generative terms, a gating network $f_{\omega}$ takes $[L_{g}^{AE}, L_d]$ as input and outputs raw weights $(u,v)$ that are normalized and lightly clipped:
\begin{equation}
(\alpha,\beta) \;=\; \operatorname{clip}_{[a,b]}\!\big(\operatorname{softmax}([u,v])\big),
\end{equation}
where $\alpha,\beta$ are adaptive mixing coefficients and the interval $[a,b]$ prevents extreme dominance of either term. Throughout this work, $a=0$, $b=1$.

The final G objective is expressed as:
\begin{equation}
\begin{aligned}
L_G^{f} \;&=\; \alpha\,L_{g}^{AE} \;+\; \beta\,L_d \;-\; \gamma\,H(\alpha,\beta),\\
\end{aligned}
\end{equation}
where the entropy regularizer is defined as:
\begin{equation}
\begin{aligned}
H(\alpha,\beta) \;&=\; -\,\alpha\log\alpha \;-\; \beta\log\beta,
\end{aligned}
\end{equation}
and $\gamma>0$ controls the strength of the entropy term. The entropy regularizer prevents early collapse of the gating weights by maintaining sufficient information entropy, thereby preserving the balance between reconstruction and adversarial losses. This mechanism reduces oscillations of loss functions $L_G^{f}$ and $L_d$, yields smoother $G_{\theta}$ updates, and improves overall training stability. In practice, AE parameters are updated on $L_{r}^{AE}$, the parameters of $C_{\phi}$ are updated under WGAN-GP with multiple critic steps per iteration, and the parameters of $G_{\theta}$ are updated on $L_G^{f}$ while keeping AE and the parameters of $C_{\phi}$ fixed during that step.

\section{Experimental Setup}

\subsection{Datasets for the Proposed Model}

This study evaluates the proposed approach using three widely adopted intrusion detection datasets, namely NSL-KDD, UNSW-NB15, and CICIDS2017. All datasets are provided in tabular form, where each row corresponds to a single network flow record. 

NSL-KDD is a refined version of KDD’99 \cite{kdd_cup_1999_data_130} that removes many redundant records to provide a more realistic difficulty level. Each record contains 41 features and a class label; among these, protocol type, service, and flag are discrete fields, while the remaining features are numerical.

UNSW-NB15 is a modern dataset covering nine attack families with 49 features plus a label. The fields of proto, service, and state are discrete features, and the remaining fields are continuous features. Training and testing follow the official UNSW-NB15 split, allocating approximately 68\% and 32\% of the data, respectively.

CICIDS2017, produced by the Canadian Institute for Cybersecurity, contains labeled traffic for one benign class and eight attack types. The commonly used subset includes approximately 78 numerical features and a categorical label; numerical fields predominate, and a small number of protocol and port fields are treated as discrete with the same codebook policy as above.

For all datasets, only the officially released flow table files are used. Continuous features are normalized to the range \mbox{$[-1.0,1.0]$}. Discrete fields are converted to integer codes using codebooks built from the training set only, and test-time categories are mapped to these codes via table lookup to prevent information leakage. For UNSW-NB15 and CICIDS2017, experimental strategies follow \cite{10312801}. For NSL-KDD, data augmentation is applied only to the minority attack classes (R2L, Probe, and U2R). No synthetic samples are generated for the two dominate classes, Normal and DoS. Details of the training and testing datasets used in this paper are summarized in Table \ref{tab:datasets_labels}.

\begin{table}[!t]
  \centering
  \begin{threeparttable}
    \caption{Datasets Information}
    \label{tab:datasets_labels}
    \scriptsize
    \renewcommand{\arraystretch}{1.05}
    \setlength{\tabcolsep}{1.5pt}
    \begin{tabular*}{\columnwidth}{@{\extracolsep{\fill}} l l c c c @{}}
      \toprule
      & & \multicolumn{3}{c}{Number of samples} \\
      \cmidrule(lr){3-5}
      Dataset & Class 
      & \makecell{Original\\Training set} 
      & \makecell{Augmented\\Training set} 
      & \makecell{Original\\Test set}\\
      \midrule
      \multirow{5}{*}{NSL-KDD}
        & Normal          & 67343 &  67343 & 9711\\
        & R2L             &  995 &  10995  & 2885\\
        & Probe           & 11656  &  21656 &  2421\\
        & DoS             & 45927  &  45927 &  7460\\
        & U2R             & 52  &  10052 &  67 \\
      \midrule
      \multirow{5}{*}{UNSW-NB15}
        & Normal          & 56000 & 56000 & 37000  \\
        & DoS             & 12264  &  22264  & 4089\\
        & Reconnaissance  & 10491  &  20491 &  3496\\
        & Shellcode       & 1133  &  11133  & 378\\
        & Worms            & 130  &  10130   & 44\\
      \midrule
      \multirow{6}{*}{CICIDS2017}
        & Benign          & 105222 & 105222 & 21045 \\
        & DoS             & 21550  &  34609 &  4310\\
        & PortScan        & 10809  &  23868  & 2162\\
        & BruteForce      & 5235  &  18294   & 1047\\
        & WebAttack       & 1476  &  14535  & 295\\
        & Bot             & 857  &  13916   & 171\\
      \bottomrule
    \end{tabular*}
    \begin{tablenotes}\footnotesize
      \item \textbf{Note:} The Class represents the network traffic class contained in the dataset.  
    \end{tablenotes}
  \end{threeparttable}
\end{table}

\subsection{IDS Models Used in Experiments}

Five different machine learning models are employed as IDS classifiers in the experiments:

\begin{itemize}
\item Two-layer LSTM: Each layer contains 64 units; the first LSTM returns sequences, followed by a dropout layer with dropout rate of 0.2 and a fully connected layer with 32 neurons \cite{9908159}. 
\item Two-layer 1D CNN: Each convolutional layer uses 64 filters with a kernel size of 5. The first convolution is followed by max-pooling of size 3, and batch normalization is applied after each convolution. Flattened features feed into a fully connected layer with 16 neurons \cite{9908159}.
\item Two-layer DNN: A simple feedforward network with two dense layers consisting of 32 and 16 neurons, respectively \cite{9908159}.
\item CNN–LSTM model: Two convolutional blocks (first with 64 filters and max-pooling of 2, second with 128 filters and max-pooling of 2), each followed by batch normalization. The output is passed to an LSTM with 100 units, with dropout of 0.2 applied on the LSTM and before the classifier \cite{bamber2025hybrid}. 
\item CNN-BiLSTM model: A convolutional layer with 32 filters and max-pooling of 2, followed by batch normalization. Features are reshaped into a 32-dimensional sequence and passed to a BiLSTM with 32 units, followed by dropout of 0.2 and a fully connected layer with 25 neurons \cite{jouhari2024lightweight}.
\end{itemize}

All five IDS models operate on per-record 1D inputs, treating the feature dimension as a single-channel sequence to enable convolutional and recurrent processing. ReLU activations are used throughout unless otherwise specified. In all IDS training experiments, the number of epochs is fixed at 100. To account for variability and ensure robustness, each IDS model is independently trained 20 times despite the inherent stability of these architectures.

This work was implemented using TensorFlow v2.8, CUDA v11.3, and cuDNN v8.1 on a Linux workstation running Ubuntu 20.04.5 LTS OS with a NVIDIA GeForce RTX3090Ti GPU and an Intel Core i7-12700KF CPU.

Using the proposed generative model GMA-SAWGAN-GP, synthetic samples were generated for each dataset listed in Table \ref{tab:datasets_labels} and merged only with its corresponding training split to form an augmented training set. IDS models were then trained on these combined datasets under both binary and multi-classification settings. For binary classification, class labels were collapsed into Normal and Abnormal, whereas for multi-classification, the original attack categories were preserved. All experiments were conducted independently for each dataset, with generation, training, and evaluation performed within the same dataset without any cross-dataset mixing. For NSL-KDD and UNSW-NB15, the official test splits were used without modification, while CICIDS2017 was partitioned into an 80\% training and 20\% testing split.

\section{Results And Discussion}

The effectiveness of the proposed generative model is comprehensively evaluated on augmented datasets against baselines using the original dataset, with binary and multi-classification results reported across multiple IDS models. In addition, the proposed approach is compared with other state-of-the-art GAN-based methods. To assess generalization, the model’s ability to detect previously unseen attacks is evaluated using the LOAO method.

\begin{table*}[htbp]
    \centering
    \begin{threeparttable}
    \caption{Binary Classification Comparison of Different IDS Models with three datasets for the Proposed model}
    \label{tab2}
    \scriptsize
    \begin{tabular*}{\textwidth}{@{\extracolsep{\fill}} l l l 
        C{1.0cm} C{1.0cm}  
        C{0.8cm} C{1.0cm}  
        C{0.8cm} C{1.0cm}} 
    \toprule
    \textbf{Dataset} & \textbf{Training Data} & \textbf{IDS Model} & \textbf{Acc.(\%)} & \textbf{F1(\%)} & \multicolumn{2}{c}{\textbf{Normal}} & \multicolumn{2}{c}{\textbf{Abnormal}} \\
    \cmidrule(lr){6-7} \cmidrule(lr){8-9}
     &  &  &  &  & \textbf{Acc.(\%)} & \textbf{F1(\%)} & \textbf{Acc.(\%)} & \textbf{F1(\%)} \\
     \midrule
     
    \multirow{10}{*}{\textbf{UNSW-NB15}} 
        & \multirow{5}{*}{Original} 
          & CNN  & 81.8$\pm$4.7 & 68.8$\pm$4.4 & 89.1$\pm$6.5 & 88.8$\pm$3.3 & 48.1$\pm$8.5 & 48.7$\pm$6.2 \\
        & & DNN   & \textbf{84.1}$\pm$1.5 & 71.5$\pm$3.5 & 91.5$\pm$0.5 & 90.5$\pm$0.8 & \textbf{50.0}$\pm$8.2 & 52.6$\pm$6.3 \\
        & & LSTM  & 84.0$\pm$3.1 & 70.2$\pm$7.2 & \textbf{92.0}$\pm$1.0 & 90.4$\pm$1.7 & 46.7$\pm$15.8 & 50.0$\pm$12.7 \\
        & & CNN-BiLSTM & 80.7$\pm$0.4 & 61.9$\pm$0.4 & 91.9$\pm$0.6 & 88.7$\pm$0.3 & 29.2$\pm$1.2 & 35.0$\pm$0.8 \\
        & & CNN-LSTM & 81.0$\pm$0.9 & 62.8$\pm$1.9 & 91.9$\pm$0.8 & 88.6$\pm$0.6 & 31.0$\pm$3.5 & 36.7$\pm$3.4 \\
    \cmidrule(lr){2-9} 
        & \multirow{5}{*}{Augmented} 
          & CNN  & 89.4$\pm$2.3 & 82.3$\pm$2.5 & 92.9$\pm$2.8 & 93.5$\pm$1.2 & 73.3$\pm$6.8 & 71.2$\pm$3.9 \\
        & & DNN   & 88.3$\pm$0.8 & 80.1$\pm$1.7 & 92.5$\pm$0.5 & 92.8$\pm$0.4 & 68.5$\pm$5.1 & 67.4$\pm$3.1 \\
        & & LSTM  & 88.3$\pm$0.9 & 81.7$\pm$1.3 & 90.3$\pm$1.2 & 92.7$\pm$0.6 & \textbf{79.2}$\pm$3.9 & 70.7$\pm$2.2 \\
        & & CNN-BiLSTM & \textbf{90.0}$\pm$0.6 & 83.6$\pm$1.1 & \textbf{92.6}$\pm$0.7 & 93.8$\pm$0.4 & 77.9$\pm$3.6 & 73.4$\pm$1.9 \\
        & & CNN-LSTM & 89.7$\pm$1.1 & 83.3$\pm$1.9 & 92.5$\pm$0.7 & 93.7$\pm$0.7 & 77.1$\pm$4.4 & 72.8$\pm$3.2 \\
    \midrule

    \multirow{10}{*}{\textbf{CICIDS2017}} 
        & \multirow{5}{*}{Original} 
          & CNN  & 93.3$\pm$1.4 & 92.1$\pm$1.5 & 91.3$\pm$2.3 & 95.2$\pm$1.1 & \textbf{98.7}$\pm$1.3 & 89.1$\pm$1.9 \\
        & & DNN   & 91.2$\pm$0.6 & 89.7$\pm$0.6 & 88.6$\pm$1.2 & 93.6$\pm$0.5 & 97.8$\pm$1.2 & 85.9$\pm$0.8 \\
        & & LSTM  & 92.9$\pm$0.8 & 91.5$\pm$0.8 & 92.4$\pm$1.3 & 95.0$\pm$0.6 & 94.4$\pm$0.7 & 88.0$\pm$1.1\\
        & & CNN-BiLSTM  & 94.1$\pm$0.7 & 93.0$\pm$0.8 & 92.7$\pm$1.1 & 95.8$\pm$0.6 & 97.9$\pm$0.5 & 90.2$\pm$1.1 \\
        & & CNN-LSTM & \textbf{94.2}$\pm$0.5 & 93.0$\pm$0.8 & \textbf{92.8}$\pm$0.8 & 95.8$\pm$0.6 & 98.1$\pm$0.4 & 90.2$\pm$1.1 \\
    \cmidrule(lr){2-9}
        & \multirow{5}{*}{Augmented}
          & CNN  & 97.3$\pm$0.6 & 97.6$\pm$0.1 & 97.4$\pm$1.2 & 98.7$\pm$0.0 & 97.1$\pm$3.7 & 96.6$\pm$0.2 \\
        & & DNN   & 97.2$\pm$0.2 & 96.5$\pm$0.2 & 97.2$\pm$0.7 & 98.0$\pm$0.1 & 97.1$\pm$1.9 & 95.0$\pm$0.3 \\
        & & LSTM  & 97.1$\pm$0.2 & 96.4$\pm$0.2 & 97.2$\pm$0.6 & 98.0$\pm$0.1 & 96.8$\pm$1.6 & 94.8$\pm$0.3 \\
        & & CNN-BiLSTM  & 98.1$\pm$0.1 & 97.6$\pm$0.1 & 97.9$\pm$0.2 & 98.7$\pm$0.0 & 98.5$\pm$0.3 & 96.6$\pm$0.2 \\
        & & CNN-LSTM & \textbf{99.2}$\pm$0.2 & 99.0$\pm$0.2 & \textbf{99.1}$\pm$0.2 & 99.4$\pm$0.1 & \textbf{99.5}$\pm$0.1 & 98.5$\pm$0.3 \\
    \midrule

    \multirow{10}{*}{\textbf{NSL-KDD}} 
        & \multirow{5}{*}{Original} 
          & CNN  & \textbf{79.0}$\pm$1.3 & 78.9$\pm$1.4 & 96.6$\pm$0.6 & 79.8$\pm$1.0 & \textbf{65.6}$\pm$2.7 & 78.0$\pm$1.8 \\
        & & DNN   & 78.1$\pm$0.7 & 78.0$\pm$0.8 & 97.4$\pm$0.3 & 79.3$\pm$0.5 & 63.5$\pm$1.4 & 76.8$\pm$1.0 \\
        & & LSTM  & \textbf{79.0}$\pm$1.1 & 79.0$\pm$1.1 & 97.4$\pm$0.1 & 80.0$\pm$0.8 & 65.1$\pm$2.0 & 77.9$\pm$1.4 \\
        & & CNN-BiLSTM  & 78.0$\pm$0.6 & 78.0$\pm$0.6 & 96.5$\pm$1.9 & 79.1$\pm$0.7 & 64.0$\pm$1.4 & 76.8$\pm$0.7 \\
        & & CNN-LSTM & 76.7$\pm$0.2 & 76.6$\pm$0.2 & \textbf{97.8}$\pm$0.1 & 78.3$\pm$0.1 & 60.8$\pm$0.3 & 74.8$\pm$0.2 \\
    \cmidrule(lr){2-9}
        & \multirow{5}{*}{Augmented} 
          & CNN  & 81.8$\pm$2.3 & 81.8$\pm$2.3 & 95.8$\pm$1.5 & 82.0$\pm$1.7 & 71.2$\pm$4.7 & 81.6$\pm$2.9 \\
        & & DNN   & 79.0$\pm$0.6 & 78.9$\pm$0.7 & 97.4$\pm$0.2 & 80.0$\pm$0.5 & 65.0$\pm$1.2 & 77.9$\pm$0.5 \\
        & & LSTM  & \textbf{84.6}$\pm$1.2 & 84.6$\pm$1.1 & 96.8$\pm$0.2 & 84.5$\pm$1.0 & 75.4$\pm$2.0 & 84.8$\pm$1.3 \\
        & & CNN-BiLSTM  & 84.4$\pm$1.9 & 84.4$\pm$1.9 & 95.7$\pm$1.7 & 84.1$\pm$1.8 & \textbf{75.8}$\pm$2.9 & 84.6$\pm$2.0 \\
        & & CNN-LSTM & 82.9$\pm$0.6 & 82.9$\pm$0.6 & \textbf{97.6}$\pm$0.2 & 83.1$\pm$0.5 & 71.8$\pm$1.1 & 82.7$\pm$0.8 \\
    \bottomrule
    \end{tabular*}
    \begin{tablenotes}
        \footnotesize
        \item \textit{Note:} The best accuracy (Acc.) of the IDS models under each dataset is highlighted in bold.
    \end{tablenotes}
    \end{threeparttable}
\end{table*}

\subsection{Evaluation Metrics}

To evaluate the performance of IDS models under class imbalance, two primary metrics are reported for each dataset: overall accuracy and F1-score, presented as the mean and standard deviation over 20 independent runs. In each run, predictions are made on the test dataset, and confusion-matrix counts - True Positives (TP), True Negatives (TN), False Positives (FP), and False Negatives (FN) - are computed. Metrics are then aggregated across the runs to calculate the reported mean and standard deviation. For the multi-classification tests, in addition to the overall accuracy and F1-score, per-class F1-score and per-class accuracy are also provided to characterize class-specific behavior. The overall accuracy summarizes aggregate test performance, whereas the F1-score assigns equal weight to each class, making it more sensitive to minority attacks. Together, these metrics provide a complementary and comprehensive view of detection capability. The metrics are defined as in (\ref{formula.accuracy}) and (\ref{formula.f1}):

\begin{equation}
\text{Accuracy} = \frac{\text{TP} + \text{TN}}{\text{TP} + \text{FP} + \text{FN} + \text{TN}}
\label{formula.accuracy}
\end{equation}

\begin{equation}
\text{F1-score} = \frac{2 \times \text{TP}}{2 \times \text{TP} + \text{FP} + \text{FN}}
\label{formula.f1}
\end{equation}

For the generalization experiment with the LOAO method, AUROC and TPR@5\%FPR are reported as the primary metrics. 
Let $s(x)$  denote the classifier score, and let $x^+$ and $x^-$ represent positive and negative samples, respectively. 
AUROC is defined as the probability that a randomly chosen positive sample is ranked above a randomly chosen negative sample:
\[
\mathrm{AUROC}=\mathbb{P}\big(s(x^+) > s(x^-)\big).
\]
For a threshold $\mu$, the True Positive Rate (TPR) and False Positive Rate (FPR) are given by:
\begin{equation}
\mathrm{TPR}(\mu)=\frac{\mathrm{TP}}{\mathrm{TP}+\mathrm{FN}},
\end{equation}
\begin{equation}
\mathrm{FPR}(\mu)=\frac{\mathrm{FP}}{\mathrm{FP}+\mathrm{TN}}.
\end{equation}
TPR@5\%FPR is defined as:
\[
\mathrm{TPR@5\%\;FPR}=\mathrm{TPR}(\mu^*),\quad \text{with}\ \mathrm{FPR}(\mu^*)=0.05,
\]
where $\mu^*$ is obtained by monotonic interpolation along the Receiver Operating Characteristic (ROC) curve.

AUROC provides a threshold-independent measure of ranking quality that is relatively robust to class imbalance \cite{shao2025robust}, while TPR@5\%FPR captures detection performance at a low false positive operating point, aligning with practical IDS requirements for unknown attack detection \cite{rendel2025zerocan}.

\subsection{Binary Classification}

In the first validation experiment, binary classification was performed on the three preprocessed datasets, with all attack categories grouped into a single Abnormal class. As shown in Table \ref{tab2}, IDS models trained on the augmented datasets consistently outperformed their counterparts trained on the original datasets. On UNSW-NB15, the average accuracies of the CNN-, DNN-, LSTM-, CNN-BiLSTM-, and CNN-LSTM–based IDS models improved by 7.6\%, 4.2\%, 4.3\%, 9.3\%, and 8.7\%, respectively. For CICIDS2017, the improvements were 4.0\%, 6.0\%, 4.2\%, 4.0\%, and 5.0\% for the same IDS models. On NSL-KDD, the average accuracy gains reached 2.8\% for CNN, 0.9\% for DNN, 5.6\% for LSTM, 6.4\% for CNN-BiLSTM, and 6.2\% for CNN-LSTM.

These results indicate that, in binary classification, the proposed GMA-SAWGAN-GP consistently sharpens the decision boundary between Normal and Abnormal flows across different IDS models. Since augmentation mainly increases the diversity and coverage of attack samples while preserving the distribution of the Normal class, the observed accuracy gains primarily reflect improved detection of Abnormal flows (fewer missed attacks) rather than trivial rebalancing toward the majority class.

The largest gains are observed for LSTM-, \mbox{CNN-BiLSTM-,} and CNN-LSTM-based IDS models, suggesting that models with greater capacity to exploit sequential or cross-feature dependencies benefit most from the additional variability introduced by the synthetic samples. Moreover, performance improvements of IDS models on UNSW-NB15 and CICIDS2017 are more pronounced than on NSL-KDD, which aligns with the richer and more recent feature representations in these two datasets. This highlights that enhanced distributional coverage is particularly valuable when underlying descriptors capture fine-grained traffic behaviors.

\subsection{Multi-Classification}

\begin{table*}[htbp]
    \centering
    \begin{threeparttable}
    \caption{Multi-Classification Results of UNSW-NB15 Dataset With Different IDS Models}
    \label{tab:results_UNSW}
    \scriptsize
    \begin{tabular*}{\textwidth}{@{\extracolsep{\fill}}
        l 
        l 
        l 
        C{0.7cm} 
        C{0.9cm} 
        C{1.0cm} C{1.0cm} 
        C{0.6cm} C{1.0cm} 
        C{0.6cm} C{1.0cm} 
        C{0.6cm} C{1.0cm} 
    }
    \toprule
    \textbf{Dataset} & \textbf{Training Data} & \textbf{IDS Model} & \textbf{Acc.(\%)} & \textbf{F1(\%)} &
    \multicolumn{2}{c}{\textbf{DoS}} &
    \multicolumn{2}{c}{\textbf{Reconnaissance}} &
    \multicolumn{2}{c}{\textbf{Shellcode}} &
    \multicolumn{2}{c}{\textbf{Worms}} \\
    \cmidrule(lr){6-7} \cmidrule(lr){8-9} \cmidrule(lr){10-11} \cmidrule(lr){12-13}
     &  &  &  &  & \textbf{Acc. (\%)} & \textbf{F1(\%)} & \textbf{Acc.(\%)} & \textbf{F1(\%)} & \textbf{Acc.(\%)} & \textbf{F1(\%)} & \textbf{Acc.(\%)} & \textbf{F1(\%)} \\
    \midrule
    
    \multirow{10}{*}{\textbf{UNSW-NB15}}
        & \multirow{5}{*}{Original}
          & CNN  & \textbf{86.5}$\pm$0.8 & 51.5$\pm$1.8  & \textbf{64.6}$\pm$9.4 & 68.5$\pm$6.0 & 43.7$\pm$3.0 & 48.1$\pm$1.9 & \textbf{45.8}$\pm$4.0   & 24.5$\pm$1.3   & 21.1$\pm$4.4   & 23.6$\pm$2.6   \\
        & & DNN   & 82.5$\pm$1.3 & 43.9$\pm$3.4   & 29.2$\pm$13.6 & 38.4$\pm$13.7 & \textbf{44.6}$\pm$1.7 & 42.3$\pm$0.9 & 37.9$\pm$4.6  & 27.2$\pm$2.4  & \textbf{24.6}$\pm$2.6 & 21.0$\pm$3.1 \\
        & & LSTM  & 84.2$\pm$1.2 & 40.8$\pm$2.0   & 58.3$\pm$13.8 & 58.6$\pm$8.2 & 33.1$\pm$1.8 & 36.4$\pm$1.8 & 14.4$\pm$3.0 & 17.4$\pm$2.1 & 0.0$\pm$0.0 & 0.0$\pm$0.0 \\
        & & CNN-BiLSTM  & 81.1$\pm$0.5 & 36.1$\pm$1.7 & 19.6$\pm$1.4 & 26.2$\pm$1.5 & 33.0$\pm$1.8 & 39.7$\pm$1.9 & 18.7$\pm$5.3 & 16.8$\pm$1.6 & 4.8$\pm$4.6 & 8.5$\pm$7.8 \\
        & & CNN-LSTM & 82.6$\pm$0.5 & 44.8$\pm$1.1   & 45.7$\pm$1.8 & 48.2$\pm$1.7 & 30.2$\pm$0.3 & 40.1$\pm$0.8 & 32.7$\pm$0.7 & 19.9$\pm$0.9 & 16.4$\pm$4.3 & 25.6$\pm$5.7 \\
    \cmidrule(lr){2-13} 
        & \multirow{5}{*}{Augmented}
          & CNN  & \textbf{89.4}$\pm$0.3 & 56.1$\pm$1.2   & \textbf{74.6}$\pm$1.8 & 75.4$\pm$1.1 & \textbf{60.1}$\pm$2.3 & 58.6$\pm$1.4 & 34.8$\pm$4.6  & 29.4$\pm$2.3  & 16.6$\pm$4.6 & 22.5$\pm$5.2 \\
        & & DNN   & 85.7$\pm$1.7 & 43.1$\pm$3.6   & 35.8$\pm$13.7 & 46.6$\pm$12.4 & 46.1$\pm$6.7 & 50.2$\pm$5.0 & 14.3$\pm$2.6 & 20.7$\pm$2.7 & 3.2$\pm$3.6 & 6.0$\pm$6.5 \\
        & & LSTM  & 85.1$\pm$0.9 & 46.0$\pm$1.4   & 54.2$\pm$8.6 & 60.8$\pm$6.6 & 45.1$\pm$2.7 & 44.6$\pm$2.0 & \textbf{50.2}$\pm$3.0  & 20.5$\pm$2.0  & \textbf{26.4}$\pm$9.4 & 10.8$\pm$2.3 \\
        & & CNN-BiLSTM  & 87.6$\pm$0.8 & 50.8$\pm$2.1 & 69.7$\pm$8.7 & 70.0$\pm$5.3 & 52.8$\pm$4.3 & 53.1$\pm$3.1 & 36.9$\pm$4.3  & 23.7$\pm$3.5  & 13.0$\pm$7.0 & 13.5$\pm$5.7 \\
        & & CNN-LSTM & 86.4$\pm$0.4 & 51.7$\pm$1.0   & 73.8$\pm$1.2 & 69.8$\pm$1.8 & 35.1$\pm$2.6 & 44.0$\pm$2.9 & 36.9$\pm$3.2 & 21.3$\pm$1.3 & 20.0$\pm$5.3 & 30.9$\pm$6.0 \\
    \bottomrule
    \end{tabular*}
    \begin{tablenotes}
        \footnotesize
        \item \textit{Note:} The best accuracy (Acc.) of the IDS models is highlighted in bold.
    \end{tablenotes}
    \end{threeparttable}
\end{table*}

\begin{table*}[htbp]
    \centering
    \begin{threeparttable}
    \caption{Multi-Classification Results of NSL-KDD Dataset With Different IDS Models}
    \label{tab:NSL_results}
    \scriptsize
    \begin{tabular*}{\textwidth}{@{\extracolsep{\fill}}
        l 
        l 
        l 
        C{0.7cm} 
        C{0.9cm} 
        C{0.6cm} C{1.0cm} 
        C{0.6cm} C{1.0cm} 
        C{0.6cm} C{1.0cm} 
        C{0.6cm} C{1.0cm} 
    }
    \toprule
    \textbf{Dataset} & \textbf{Training Data} & \textbf{IDS Model} & \textbf{Acc.(\%)} & \textbf{F1(\%)} &
    \multicolumn{2}{c}{\textbf{R2L}} &
    \multicolumn{2}{c}{\textbf{Probe}} &
    \multicolumn{2}{c}{\textbf{DoS}} &
    \multicolumn{2}{c}{\textbf{U2R}} \\
    \cmidrule(lr){6-7} \cmidrule(lr){8-9} \cmidrule(lr){10-11} \cmidrule(lr){12-13}
     &  &  &  &  & \textbf{Acc.(\%)} & \textbf{F1(\%)} & \textbf{Acc.(\%)} & \textbf{F1(\%)} & \textbf{Acc.(\%)} & \textbf{F1(\%)} & \textbf{Acc.(\%)} & \textbf{F1(\%)} \\
    \midrule
    
    \multirow{10}{*}{\textbf{NSL-KDD}}
        & \multirow{5}{*}{Original}
          & CNN  & 76.8$\pm$0.3 & 56.6$\pm$1.4  & 1.8$\pm$0.6 & 3.5$\pm$1.1 & \textbf{77.7}$\pm$2.8 & 78.5$\pm$1.7 & 80.0$\pm$1.5   & 87.6$\pm$0.9   & 21.1$\pm$4.2   & 33.8$\pm$5.6   \\
        & & DNN   & 74.4$\pm$0.6 & 51.9$\pm$1.8   & 0.6$\pm$0.5 & 1.2$\pm$1.0 & 59.2$\pm$3.6 & 67.6$\pm$2.4 & 77.9$\pm$0.9  & 86.1$\pm$0.5  & 15.9$\pm$5.2 & 26.8$\pm$7.7 \\
        & & LSTM  & 75.7$\pm$0.5 & 55.8$\pm$1.0   & 2.7$\pm$2.2 & 5.1$\pm$4.0 & 74.4$\pm$2.4 & 72.0$\pm$2.1 & 76.8$\pm$1.0 & 85.5$\pm$0.7 & \textbf{24.9}$\pm$4.4 & 36.4$\pm$5.0 \\
        & & CNN-BiLSTM  & 75.1$\pm$1.5 & 53.5$\pm$2.5 & 4.2$\pm$2.3 & 7.9$\pm$4.2 & 58.8$\pm$4.5 & 65.4$\pm$4.5 & 79.1$\pm$3.4 & 86.8$\pm$2.1 & 18.4$\pm$7.0 & 28.6$\pm$8.4 \\
        & & CNN-LSTM & \textbf{77.0}$\pm$0.6 & 55.3$\pm$1.0   & \textbf{6.2}$\pm$1.2 & 11.6$\pm$2.1 & 54.4$\pm$4.1 & 65.9$\pm$3.6 & \textbf{85.6}$\pm$0.2 & 90.6$\pm$0.2 & 17.9$\pm$3.5 & 29.1$\pm$4.9 \\
    \cmidrule(lr){2-13} 
        & \multirow{5}{*}{Augmented}
          & CNN  & 78.8$\pm$0.7 & 59.5$\pm$2.1   & 3.8$\pm$2.1 & 7.2$\pm$3.8 & \textbf{81.2}$\pm$1.3 & 81.5$\pm$0.9 & 84.4$\pm$1.5  & 90.2$\pm$0.8  & 24.6$\pm$5.3 & 38.1$\pm$6.8 \\
        & & DNN   & 74.2$\pm$0.2 & 54.5$\pm$1.0   & 0.8$\pm$0.5 & 1.5$\pm$1.0 & 61.2$\pm$3.6 & 69.2$\pm$2.4 & 76.6$\pm$1.0 & 85.4$\pm$0.6 & \textbf{26.1}$\pm$3.1 & 38.8$\pm$4.1 \\
        & & LSTM  & 79.6$\pm$0.8 & 57.0$\pm$2.4   & 16.9$\pm$3.5 & 28.4$\pm$5.2 & 78.7$\pm$2.8 & 75.7$\pm$4.3 & 82.0$\pm$1.6  & 88.7$\pm$1.0  & 5.5$\pm$6.9 & 8.9$\pm$11.0 \\
        & & CNN-BiLSTM  & 79.8$\pm$1.2 & 59.2$\pm$2.9 & 12.5$\pm$4.5 & 21.9$\pm$7.4 & 69.9$\pm$6.2 & 75.3$\pm$4.5 & 86.7$\pm$1.6  & 90.2$\pm$1.5  & 16.7$\pm$4.9 & 25.9$\pm$6.3 \\
        & & CNN-LSTM & \textbf{81.4}$\pm$0.7 & 62.6$\pm$1.5   & \textbf{19.3}$\pm$4.4 & 32.0$\pm$6.3 & 71.6$\pm$3.2 & 78.0$\pm$2.1 & \textbf{88.8}$\pm$1.0 & 92.6$\pm$0.6 & 16.9$\pm$3.1 & 27.6$\pm$4.3 \\
    \bottomrule
    \end{tabular*}
    \begin{tablenotes}
        \footnotesize
        \item \textit{Note:} The best accuracy (Acc.) of the IDS models is highlighted in bold.
    \end{tablenotes}
    \end{threeparttable}
\end{table*}

\begin{table*}[htbp]
    \centering
    \begin{threeparttable}
    \caption{Multi-Classification Results of CICIDS2017 Dataset With Different IDS Models}
    \label{tab:CICIDS_results}
    \tiny
    \begin{tabular*}{\textwidth}{@{\extracolsep{\fill}}
        l 
        l 
        l 
        C{0.6cm} 
        C{0.6cm} 
        C{0.6cm} C{0.6cm} 
        C{0.6cm} C{0.6cm} 
        C{0.6cm} C{0.6cm} 
        C{0.6cm} C{0.6cm} 
        C{0.6cm} C{0.7cm} 
    }
    \toprule
    \textbf{Dataset} & \textbf{Training Data} & \textbf{IDS Model} & \textbf{Acc.(\%)} & \textbf{F1(\%)} &
    \multicolumn{2}{c}{\textbf{DoS}} &
    \multicolumn{2}{c}{\textbf{PortScan}} &
    \multicolumn{2}{c}{\textbf{BruteForce}} &
    \multicolumn{2}{c}{\textbf{WebAttack}} &
    \multicolumn{2}{c}{\textbf{Bot}} \\  %
    \cmidrule(lr){6-7} \cmidrule(lr){8-9} \cmidrule(lr){10-11} \cmidrule(lr){12-13} \cmidrule(lr){14-15}
     &  &  &  &  & \textbf{Acc.(\%)} & \textbf{F1(\%)} & \textbf{Acc.(\%)} & \textbf{F1(\%)} & \textbf{Acc.(\%)} & \textbf{F1(\%)} & \textbf{Acc.(\%)} & \textbf{F1(\%)} & \textbf{Acc.(\%)} & \textbf{F1(\%)} \\
    \midrule
    
    \multirow{10}{*}{\textbf{CICIDS2017}}
        & \multirow{5}{*}{Original}
          & CNN      & 97.3$\pm$0.2 & 92.8$\pm$1.5  & 98.3$\pm$0.9 & 96.7$\pm$0.4 & \textbf{99.8}$\pm$0.1 & 93.4$\pm$0.2 & 99.1$\pm$0.3 & 97.4$\pm$0.6 & 91.9$\pm$1.1 & 83.4$\pm$0.7 & 90.6$\pm$11.9 & 88.1$\pm$8.0 \\
        & & DNN      & 96.8$\pm$0.2 & 89.3$\pm$2.5   & 93.9$\pm$2.5 & 95.5$\pm$0.6 & 98.3$\pm$1.8 & 93.2$\pm$0.3 & 98.6$\pm$0.7 & 97.6$\pm$0.4 & 85.7$\pm$1.0 & 80.7$\pm$1.1 & 60.5$\pm$16.6 & 71.1$\pm$13.6 \\
        & & LSTM     & 96.5$\pm$0.3 & 90.0$\pm$0.8   & 97.0$\pm$1.1 & 95.8$\pm$0.3 & 99.3$\pm$0.2 & 93.0$\pm$0.1 & 98.0$\pm$2.1 & 92.0$\pm$2.7 & 88.7$\pm$1.0 & 81.9$\pm$0.6 & 80.4$\pm$3.3 & 79.4$\pm$2.6 \\
        & & CNN-BiLSTM & 98.1$\pm$0.2 & 94.9$\pm$0.5 & 98.9$\pm$0.3 & 98.0$\pm$0.3 & 97.6$\pm$1.5 & 94.1$\pm$0.6 & 99.0$\pm$0.4 & 98.6$\pm$0.3 & 93.9$\pm$1.2 & 84.5$\pm$1.8 & \textbf{98.8}$\pm$0.3 & 95.5$\pm$0.9 \\
        & & CNN-LSTM & \textbf{98.4}$\pm$0.0 & 95.5$\pm$0.1   & \textbf{99.5}$\pm$0.1 & 98.5$\pm$0.0 & 97.1$\pm$0.7 & 94.6$\pm$0.1 & \textbf{99.2}$\pm$0.1 & 99.1$\pm$0.0 & \textbf{94.3}$\pm$0.4 & 85.4$\pm$0.5 & \textbf{98.8}$\pm$0.2 & 96.9$\pm$0.4 \\
    \cmidrule(lr){2-15} 
        & \multirow{5}{*}{Augmented}
          & CNN      & 97.4$\pm$0.4 & 92.5$\pm$2.2   & 98.2$\pm$0.8 & 96.8$\pm$0.5 & \textbf{99.4}$\pm$0.6 & 93.8$\pm$0.5 & \textbf{99.2}$\pm$0.3  & 96.5$\pm$1.4  & 91.1$\pm$1.2 & 83.1$\pm$1.3 & 86.8$\pm$13.1 & 86.9$\pm$9.4 \\
        & & DNN      & 96.7$\pm$0.3 & 88.6$\pm$3.1   & 92.7$\pm$1.7 & 95.1$\pm$0.6 & 98.3$\pm$1.8 & 93.1$\pm$0.4 & 98.1$\pm$0.5 & 96.9$\pm$0.4 & 86.0$\pm$1.4 & 81.1$\pm$1.0 & 55.5$\pm$19.6 & 67.4$\pm$16.9 \\
        & & LSTM     & 97.3$\pm$0.2 & 93.6$\pm$0.4   & 94.9$\pm$1.5 & 96.0$\pm$0.4 & 97.0$\pm$0.8 & 94.1$\pm$0.2 & 99.1$\pm$0.1  & 97.5$\pm$0.5  & 90.2$\pm$0.5 & 82.6$\pm$0.5 & 96.7$\pm$0.8 & 93.1$\pm$1.5 \\
        & & CNN-BiLSTM & 98.1$\pm$0.1 & 95.0$\pm$0.5 & 98.4$\pm$0.4 & 98.0$\pm$0.2 & 97.8$\pm$0.8 & 94.3$\pm$0.3 & 98.8$\pm$0.4  & 98.3$\pm$0.3  & 92.3$\pm$1.5 & 84.9$\pm$1.4 & 98.4$\pm$0.7 & 95.8$\pm$1.0 \\
        & & CNN-LSTM & \textbf{98.4}$\pm$0.0 & 95.6$\pm$0.1   & \textbf{99.5}$\pm$0.1 & 98.5$\pm$0.1 & 97.4$\pm$0.2 & 94.6$\pm$0.1 & 99.0$\pm$0.1 & 99.0$\pm$0.1 & \textbf{94.2}$\pm$0.4 & 85.6$\pm$0.7 & \textbf{98.8}$\pm$0.0 & 97.1$\pm$0.1 \\
    \bottomrule
    \end{tabular*}
    \begin{tablenotes}
        \footnotesize
        \item \textit{Note:} The best accuracy (Acc.) of the IDS models is highlighted in bold.
    \end{tablenotes}
    \end{threeparttable}
\end{table*}

Table \ref{tab:results_UNSW} presents the multi-classification results on UNSW-NB15. For overall average accuracy, all IDS models trained on the augmented dataset outperform their corresponding IDS models trained on the original dataset. The CNN-BiLSTM-based IDS model achieves the highest performance improvement at 6.5\%. Consistent gains are observed for the DoS and Reconnaissance classes, where additional synthetic traffic reinforces the decision boundary between these attacks and benign flows. In contrast, for the Shellcode and Worms classes, accuracy improves for LSTM-, CNN-LSTM-, and CNN-BiLSTM-based IDS models, while CNN- and DNN-based IDS models show slight reductions. This behavior reflects the ability of recurrent architectures to exploit non-local, cross-feature dependencies introduced by synthetic samples, whereas CNN- and DNN-based IDS models rely on short receptive fields or global dense layers and are more sensitive to subtle distribution shifts, making them less robust for minority attack classes with very few samples \cite{zhuo2022cross}.

Thus, on UNSW-NB15, GMA-SAWGAN-GP mainly strengthens majority-class decision boundaries for all IDS models, while its benefits on rare Shellcode and Worms classes are concentrated in LSTM-, CNN-LSTM-, and CNN-BiLSTM-based IDS models that can exploit long-range, cross-feature dependencies to reduce false negatives on these attacks.

Table \ref{tab:NSL_results} reports the multi-classification results on NSL-KDD. Average accuracy has increased by 2.0\%, 3.9\%, 4.7\%, and 4.4\% for CNN-, LSTM-, CNN-BiLSTM-, and CNN-LSTM-based IDS models, respectively, whereas the DNN-based IDS model shows weak overall performance. This underperformance stems from the lack of inductive bias for cross-feature dependencies in tabular flows, making the DNN-based IDS model prone to overfitting under class imbalance. Additionally, its reliance on BatchNorm and specific optimization settings increases sensitivity to moment drift, reducing generalization. NSL-KDD further exacerbates these limitations due to extreme minority sparsity for classes R2L and U2R and legacy, partially redundant features \cite{abbasi2024enhanced}, which amplify class overlap and hinder robust learning for the DNN model. 

This experiment suggests that, on NSL-KDD, generative augmentation primarily helps CNN-, LSTM-, \mbox{CNN-LSTM-,} and CNN-BiLSTM-based IDS models to regularize decision boundaries under extreme minority sparsity, improving the separability of R2L and U2R classes from Normal and DoS classes. In contrast, the DNN-based IDS model remains limited by feature redundancy and heightened sensitivity to distributional drift.

Table \ref{tab:CICIDS_results} shows the multi-classification results on CICIDS2017. After incorporating synthetic data, hybrid IDS models maintained an overall accuracy near saturation, with little changes in accuracies and F1-scores for CNN-BiLSTM- and CNN-LSTM-based IDS models. The LSTM-based IDS model benefited most, with average accuracy and macro F1-score increasing by 0.8\% and 3.6\%, respectively, while the CNN- and DNN-based IDS models remained largely unchanged or declined slightly. DoS and BruteForce classes were already near their maximum potential, yielding negligible improvements. Gains on PortScan and Bot classes were concentrated in the LSTM-based IDS model, whereas CNN- and DNN-based IDS models exhibited small declines on Bot class, indicating greater sensitivity to distribution shifts introduced by augmentation. For the LSTM-based IDS model, synthetic samples improved detection reliability for PortScan and Bot and yielded consistent F1-score gains across other attack classes. This advantage reflects the model’s ability to exploit the richer cross-feature interaction patterns introduced by the generator G, particularly for underrepresented and structurally complex attacks. In contrast, the CNN-LSTM-based IDS model operated near a performance ceiling on the original dataset, with high accuracy and F1-scores across most attack classes; thus, additional synthetic samples were largely redundant, only marginally adjusting the decision boundaries and producing negligible improvements under augmentation.

\subsection{Generalization}

The proposed generative model's generalizability to unknown attacks is evaluated using the LOAO method on NSL-KDD, UNSW-NB15, and CICIDS2017. For each dataset, two attack classes are designated as unknown: Probe and R2L for NSL-KDD, DoS and Reconnaissance for UNSW-NB15, and DoS and PortScan for CICIDS2017. In each LOAO trial, the designated class is completely excluded from the training set and reserved for testing, while the remaining attack classes and normal traffic constitute the training set. The test set remains fixed across all conditions. Detection performance is quantified using AUROC and TPR@5\%FPR, reported as the mean with standard deviation over 20 independent runs. Models trained on the original dataset are compared with models trained on the augmented dataset that includes synthetic samples generated by GMA-SAWGAN-GP. This setup evaluates whether the proposed novel generative augmentation improves IDS capability to detect previously unknown attacks.

\begin{table*}[t]
\centering
\caption{LOAO Results Across Five IDS Models on NSL-KDD, UNSW-NB15, and CICIDS2017}
\label{tab:loao_five_models}
\renewcommand{\arraystretch}{1.08}
\setlength{\tabcolsep}{2pt} %
\scriptsize

\begin{tabular}{l c l c c c c}
\toprule
\multirow{2}{*}{Dataset} & \multirow{2}{*}{Unknown Attack} & \multirow{2}{*}{Model}
& \multicolumn{2}{c}{Original Dataset} & \multicolumn{2}{c}{Augmented Dataset} \\
\cmidrule(lr){4-5}\cmidrule(lr){6-7}
 &  &  & AUROC (\%) & TPR@5\%FPR (\%) & AUROC (\%)  & TPR@5\%FPR (\%) \\
\midrule
\multirow{10}{*}{NSL–KDD}
& \multirow{5}{*}{Probe}
& CNN & 84.4$\pm$2.9 & 53.3$\pm$1.0 & 85.6$\pm$1.2 & 54.6$\pm$8.8  \\
& & DNN & 68.3$\pm$11.5 & 51.4$\pm$4.7 & 75.7$\pm$9.4 & 72.5$\pm$13.2 \\
& & LSTM & 72.5$\pm$4.7 & 44.9$\pm$11.0 & 76.4$\pm$2.5 & 39.5$\pm$10.3\\
& & CNN-BiLSTM & 81.3$\pm$4.4& 53.5$\pm$12.2 & 83.1$\pm$2.0 & 56.3$\pm$10.1 \\
& & CNN-LSTM & 78.4$\pm$4.1 & 31.0$\pm$12.0 & 79.5$\pm$3.4 & 33.5$\pm$11.1 \\
\cmidrule(lr){2-7}
& \multirow{5}{*}{R2L}
& CNN & 76.9$\pm$2.5 & 21.2$\pm$7.4 & 76.6$\pm$2.1 & 26.9$\pm$6.2 \\
& & DNN & 55.1$\pm$5.6 & 22.3$\pm$8.0 & 60.0$\pm$7.2 & 28.4$\pm$10.5\\
& & LSTM & 65.1$\pm$6.3 & 11.6$\pm$4.8 & 57.4$\pm$5.2 & 10.1$\pm$2.7 \\
& & CNN-BiLSTM & 71.8$\pm$4.5 & 28.0$\pm$7.1 & 72.7$\pm$2.5 & 30.0$\pm$4.3 \\
& & CNN-LSTM & 78.2$\pm$2.0 & 36.0$\pm$5.2 & 78.4$\pm$2.0 & 33.8$\pm$4.7  \\
\midrule
\multirow{10}{*}{UNSW–NB15}
& \multirow{5}{*}{DoS}
& CNN & 73.2$\pm$7.5 & 26.5$\pm$10.0 & 75.9$\pm$1.7 & 16.5$\pm$3.4 \\
& & DNN & 56.7$\pm$17.4 & 11.9$\pm$7.5 & 71.0$\pm$4.7 & 17.2$\pm$10.6 \\
& & LSTM & 76.7$\pm$2.6 & 5.3$\pm$10.6 & 66.5$\pm$2.0 & 28.9$\pm$1.4 \\
& & CNN-BiLSTM & 54.6$\pm$4.7 & 8.6$\pm$1.8 & 75.1$\pm$2.8 & 16.3$\pm$6.3\\
& & CNN-LSTM & 54.2$\pm$4.1 & 6.7$\pm$1.8 & 73.5$\pm$5.3 & 12.8$\pm$8.0 \\
\cmidrule(lr){2-7}
& \multirow{5}{*}{Reconnaissance}
& CNN & 84.0$\pm$5.5 & 40.9$\pm$7.6 & 85.6$\pm$1.8 & 55.7$\pm$5.9 \\
& & DNN & 67.0$\pm$16.2 & 16.0$\pm$9.0 & 77.7$\pm$5.7 & 36.4$\pm$4.7 \\
& & LSTM & 75.6$\pm$3.3 & 28.0$\pm$9.8 & 74.2$\pm$2.4 & 15.7$\pm$3.5 \\
& & CNN-BiLSTM & 67.9$\pm$3.2 & 20.7$\pm$2.1 & 78.5$\pm$2.9 & 34.8$\pm$5.7 \\
& & CNN-LSTM & 70.7$\pm$2.0 & 23.1$\pm$4.3 & 80.9$\pm$3.5 & 33.4$\pm$6.9 \\
\midrule
\multirow{10}{*}{CICIDS2017}
& \multirow{5}{*}{DoS}
& CNN & 50.6$\pm$6.7  & 5.7$\pm$5.5 & 59.3$\pm$6.8 & 13.3$\pm$4.1 \\
& & DNN & 34.8$\pm$13.8 & 2.0$\pm$3.3 & 29.1$\pm$15.0 & 0.2$\pm$0.4 \\
& & LSTM & 57.0$\pm$6.2 & 2.0$\pm$1.5 & 72.7$\pm$5.9 & 12.4$\pm$5.3 \\
& & CNN-BiLSTM & 60.5$\pm$12.8 & 0.6$\pm$1.1 & 79.6$\pm$4.7 & 10.4$\pm$5.8 \\
& & CNN-LSTM & 67.1$\pm$7.4 & 1.6$\pm$2.4 & 72.6$\pm$5.5 & 2.6$\pm$3.1 \\
\cmidrule(lr){2-7}
& \multirow{5}{*}{PortScan}
& CNN & 83.9$\pm$2.0 & 13.3$\pm$8.3  & 81.8$\pm$2.8 & 20.0$\pm$8.5 \\
& & DNN & 81.0$\pm$11.1 & 17.6$\pm$13.8 & 53.0$\pm$16.2 & 4.0$\pm$9.7 \\
& & LSTM & 78.2$\pm$6.1 & 3.3$\pm$4.5 & 84.2$\pm$2.6 & 8.6$\pm$6.0 \\
& & CNN-BiLSTM & 78.1$\pm$3.6 & 10.2$\pm$6.5 & 83.0$\pm$6.4 & 18.6$\pm$13.6  \\
& & CNN-LSTM & 79.4$\pm$3.3 & 13.9$\pm$13.1 & 81.2$\pm$3.5 & 12.3$\pm$8.6 \\
\bottomrule
\end{tabular}
\end{table*}

Table \ref{tab:loao_five_models} summarizes the LOAO results across three datasets and five IDS models, demonstrating that incorporating generated data enhances generalization to unknown attacks. Compared to baselines trained solely on original data, models trained on the augmented dataset (original + synthetic) achieve higher AUROC on unknown attacks and achieve substantive gains in TPR@5\%FPR overall. For instance, on NSL-KDD with Probe as the unknown attack type, the CNN-based IDS model improves AUROC by 1.2\%, reaching 85.6\%. On UNSW-NB15 with Reconnaissance as the unknown attack type, CNN-, DNN-, CNN-BiLSTM-, and CNN-LSTM-based IDS models all show AUROC improvements, with the CNN-based IDS model achieving 85.6\%. Similarly, on CICIDS2017 with PortScan as the unknown attack type, the LSTM-based IDS model records a 6\% AUROC increase, reaching 84.2\%. 

To avoid conflating ranking improvements with deployable gains, TPR@5\%FPR is also reported. Under this evaluation, training with generated data produces consistent benefits across most IDS models and is accompanied by lower standard deviations, indicating that generalization manifests not only in higher mean performance but also in more stable outcomes. A small subset of IDS models exhibits declines in TPR@5\%FPR, for instance, on UNSW-NB15 with DoS as the unknown attack type, the LSTM-based IDS model shows a decrease, attributed to class-conditional mismatch between synthetic and real distributions. We also observe that when an attack class with a relatively small number of samples is designated unknown, improvements in AUROC and TPR@5\%FPR are modest or may even decline. This suggests that generative augmentation primarily operates within known attack classes, aligning the decision boundary closely with their distributions. When the unknown class is both sparse and overlaps strongly in feature space with known attack or normal traffic, its samples are less likely to fall within the expanded attack subspace, resulting in insufficient score gains in TPR@5\%FPR. Furthermore, relatively large standard deviations arise because the unknown attack classes contain very few samples, making AUROC and especially TPR@5\%FPR highly sensitive to random initialization and mini-batch composition across runs.

Overall, training with generated data effectively expands coverage of unknown patterns and enhances transferability and robustness for detecting previously unknown attacks. However, while generative augmentation improves generalization of IDS models to unknown attacks, the small sample size of unknown attacks introduces higher variability, offsetting potential mean gains and limiting improvements for a small proportion of attacks under the LOAO evaluation method.

\newcolumntype{C}{>{\centering\arraybackslash}m{1.4cm}} %
\setlength{\tabcolsep}{2pt}
\renewcommand{\arraystretch}{1.08}
\begin{table*}[htbp]
\centering
\caption{Comparison of classification accuracies between the proposed method and State-of-the-art GAN models}
\label{tab:cmp_by_ids}
  \begin{adjustbox}{center, max width=\textwidth}
  \begin{threeparttable}
  \scriptsize             
    \begin{tabular}{l l C C C C C C}
      \toprule
      \multirow{2}{*}{IDS model} & \multirow{2}{*}{GAN Models}
      & \multicolumn{2}{c}{UNSW\textendash NB15}
      & \multicolumn{2}{c}{NSL\textendash KDD}
      & \multicolumn{2}{c}{CICIDS2017} \\
      \cmidrule(lr){3-4}\cmidrule(lr){5-6}\cmidrule(lr){7-8}
      & & B.Acc (\%) & M.Acc (\%) & B.Acc (\%) & M.Acc (\%) & B.Acc (\%) & M.Acc (\%) \\
      \midrule
      \multirow{6}{*}{CNN}
      & SYN-GAN \cite{rahman2024syn}      & 81.1$\pm$2.6 & 86.6$\pm$0.9 & 80.4$\pm$1.3 & 76.7$\pm$0.6 & 95.5$\pm$2.6 & \textbf{97.8}$\pm$0.0 \\
      & DAE-GAN \cite{9893038}     & 85.6$\pm$0.9 & 88.6$\pm$1.1 & 77.5$\pm$1.2 & 77.2$\pm$0.6 & 94.7$\pm$0.5 & \underline{97.6}$\pm$0.1 \\
      & MultiCritics-WGAN-GP \cite{10852156} & 83.1$\pm$2.1 & \underline{88.7}$\pm$1.2 & 81.1$\pm$2.6 & 77.4$\pm$0.5 & \underline{96.8}$\pm$0.8 & 97.3$\pm$0.0 \\
      & TMG-GAN \cite{10312801}      & 82.1$\pm$2.5 & 88.1$\pm$0.6 & 79.1$\pm$1.6 & 77.3$\pm$0.7 & 95.3$\pm$3.7 & 96.8$\pm$0.5 \\
      & WCGAN-GP \cite{srivastava2023wcgan}     & \underline{87.9}$\pm$2.3 & 87.3$\pm$0.5 & \textbf{83.7}$\pm$1.6 & \textbf{79.4}$\pm$0.8 & \textbf{97.3}$\pm$0.7 & 97.3$\pm$0.2 \\
      & The Proposed GMA-SAWGAN-GP    & \textbf{89.4}$\pm$2.3 & \textbf{89.4}$\pm$0.3 & \underline{81.8}$\pm$2.3 & \underline{78.8}$\pm$0.7 & \textbf{97.3}$\pm$0.6 & 97.4$\pm$0.4 \\
      \midrule
      \multirow{6}{*}{DNN}
      & SYN-GAN \cite{rahman2024syn}      & 83.1$\pm$0.4 & 83.3$\pm$0.5 & 77.8$\pm$0.4 & 75.0$\pm$0.7 & \underline{97.1}$\pm$0.3 & \textbf{96.8}$\pm$0.3 \\
      & DAE-GAN \cite{9893038}      & \underline{89.2}$\pm$0.9 & \textbf{87.8}$\pm$1.0 & \textbf{81.4}$\pm$0.7 & \underline{75.7}$\pm$0.9 & 96.6$\pm$0.1 & 96.6$\pm$0.2 \\
      & MultiCritics-WGAN-GP \cite{10852156} & 83.1$\pm$0.4 & 83.3$\pm$0.5 & 79.6$\pm$1.4 & 75.3$\pm$0.9 & 96.7$\pm$0.4 & 96.0$\pm$0.2 \\
      & TMG-GAN \cite{10312801}      & 84.5$\pm$1.1 & 85.2$\pm$1.3 & 80.1$\pm$2.1 & 75.1$\pm$0.7 & 96.8$\pm$0.3 & 95.4$\pm$0.8 \\
      & WCGAN-GP \cite{srivastava2023wcgan}     & \textbf{89.9}$\pm$0.9 & \underline{87.5}$\pm$0.6 & \underline{80.4}$\pm$1.1 & \textbf{77.1}$\pm$0.6 & 95.4$\pm$1.4 & 95.0$\pm$0.4 \\
      & The Proposed GMA-SAWGAN-GP     & 88.3$\pm$0.8 & 85.7$\pm$1.7 & 79.0$\pm$0.6 & 74.2$\pm$0.2 & \textbf{97.2}$\pm$0.2 & \underline{96.7}$\pm$0.3 \\
      \midrule
      \multirow{6}{*}{LSTM}
      & SYN-GAN \cite{rahman2024syn}      & 84.6$\pm$2.5 & 84.9$\pm$0.8 & 78.4$\pm$1.0 & 75.4$\pm$0.6 & \underline{97.0}$\pm$0.2 & \textbf{97.3}$\pm$0.1 \\
      & DAE-GAN \cite{9893038}      & 85.0$\pm$1.5 & 83.9$\pm$1.0 & 77.7$\pm$1.4 & 76.7$\pm$0.8 & 96.3$\pm$0.2 & 96.8$\pm$0.2 \\
      & MultiCritics-WGAN-GP \cite{10852156} & 83.3$\pm$2.1 & 82.4$\pm$1.1 & 79.8$\pm$1.2 & \underline{77.1}$\pm$1.2 & 96.3$\pm$0.5 & \underline{96.9}$\pm$0.3 \\
      & TMG-GAN \cite{10312801}      & 82.9$\pm$1.7 & \underline{85.1}$\pm$0.9 & \underline{81.5}$\pm$0.7 & 76.3$\pm$0.4 & 96.4$\pm$0.2 & \underline{96.9}$\pm$0.2 \\
      & WCGAN-GP \cite{srivastava2023wcgan}     & \textbf{90.6}$\pm$0.5 & \textbf{87.5}$\pm$1.2 & 80.0$\pm$2.2 & 76.7$\pm$1.1 & 96.9$\pm$0.1 & \textbf{97.3}$\pm$0.1 \\
      & The Proposed GMA-SAWGAN-GP     & \underline{88.3}$\pm$0.9 & \underline{85.1}$\pm$0.9 & \textbf{84.6}$\pm$1.2 & \textbf{79.6}$\pm$0.8 & \textbf{97.1}$\pm$0.2 & \textbf{97.3}$\pm$0.2 \\
      \midrule
      \multirow{6}{*}{CNN-BiLSTM}
      & SYN-GAN \cite{rahman2024syn}     & 81.7$\pm$0.4 & 81.7$\pm$0.6 & 77.7$\pm$1.6 & 76.0$\pm$1.5 & \textbf{98.5}$\pm$0.2 & 98.2$\pm$0.2 \\
      & DAE-GAN \cite{9893038}      & 82.8$\pm$0.8 & 82.2$\pm$1.3 & 78.6$\pm$2.2 & 75.6$\pm$0.5 & 98.2$\pm$0.2 & 98.2$\pm$0.2 \\
      & MultiCritics-WGAN-GP \cite{10852156} & 81.4$\pm$1.5 & 81.1$\pm$0.8 & 80.6$\pm$1.5 & 76.7$\pm$1.8 & 97.6$\pm$0.5 & \underline{98.4}$\pm$0.5 \\
      & TMG-GAN \cite{10312801}      & 82.4$\pm$2.6 & \underline{83.0}$\pm$1.4 & 80.0$\pm$2.3 & 77.7$\pm$2.1 & \underline{98.4}$\pm$0.5 & \textbf{98.6}$\pm$0.3 \\
      & WCGAN-GP \cite{srivastava2023wcgan}     & \underline{88.2}$\pm$0.6 & 82.6$\pm$1.7 & \textbf{85.4}$\pm$2.3 & \textbf{81.2}$\pm$1.2 & 98.1$\pm$0.6 & 98.0$\pm$0.2 \\
      & The Proposed GMA-SAWGAN-GP     & \textbf{90.0}$\pm$0.6 & \textbf{87.6}$\pm$0.8 & \underline{84.4}$\pm$1.9 & \underline{79.8}$\pm$1.2 & 98.1$\pm$0.1 & 98.1$\pm$0.1 \\
      \midrule
      \multirow{6}{*}{CNN-LSTM}
      & SYN-GAN \cite{rahman2024syn}      & 81.9$\pm$0.6 & 82.8$\pm$0.4 & 76.4$\pm$0.8 & 77.8$\pm$0.8 & 98.5$\pm$0.3 & \textbf{98.4}$\pm$0.1 \\
      & DAE-GAN \cite{9893038}     & 85.7$\pm$1.4 & 82.0$\pm$0.4 & 78.6$\pm$2.2 & 75.6$\pm$0.5 & 98.3$\pm$0.2 & 98.2$\pm$0.0 \\
      & MultiCritics-WGAN-GP \cite{10852156} & 84.3$\pm$1.6 & 85.6$\pm$0.8 & 77.6$\pm$1.0 & 79.5$\pm$0.6 & 98.5$\pm$0.4 & 98.2$\pm$0.2 \\
      & TMG-GAN \cite{10312801}      & 86.5$\pm$0.9 & \textbf{87.5}$\pm$0.4 & 79.0$\pm$2.0 & 81.3$\pm$0.5 & 98.6$\pm$0.4 & \underline{98.3}$\pm$0.0 \\
      & WCGAN-GP \cite{srivastava2023wcgan}     & \underline{89.4}$\pm$0.9 & 86.1$\pm$0.7 & \underline{82.6}$\pm$0.7 & \textbf{82.1}$\pm$0.7 & \underline{99.1}$\pm$0.4 & \underline{98.3}$\pm$0.0 \\
      & The Proposed GMA-SAWGAN-GP     & \textbf{89.7}$\pm$1.1 & \underline{86.4}$\pm$0.4 & \textbf{82.9}$\pm$0.6 & \underline{81.4}$\pm$0.7 & \textbf{99.2}$\pm$0.2 & \textbf{98.4}$\pm$0.0 \\
      \bottomrule
    \end{tabular}

  \begin{tablenotes}
    \scriptsize
    \item \textit{Note:} Results are Binary classification Accuracy (B.Acc) and Multi-classification Accuracy (M.Acc). The best average accuracy for each IDS model is highlighted in bold, and the second-best value is underlined.
  \end{tablenotes}
\end{threeparttable}
\end{adjustbox}
\end{table*}

\subsection{Comparison with Related Works}

\begin{figure*}[htbp]
  \centering
  \includegraphics[width=0.75\textwidth]{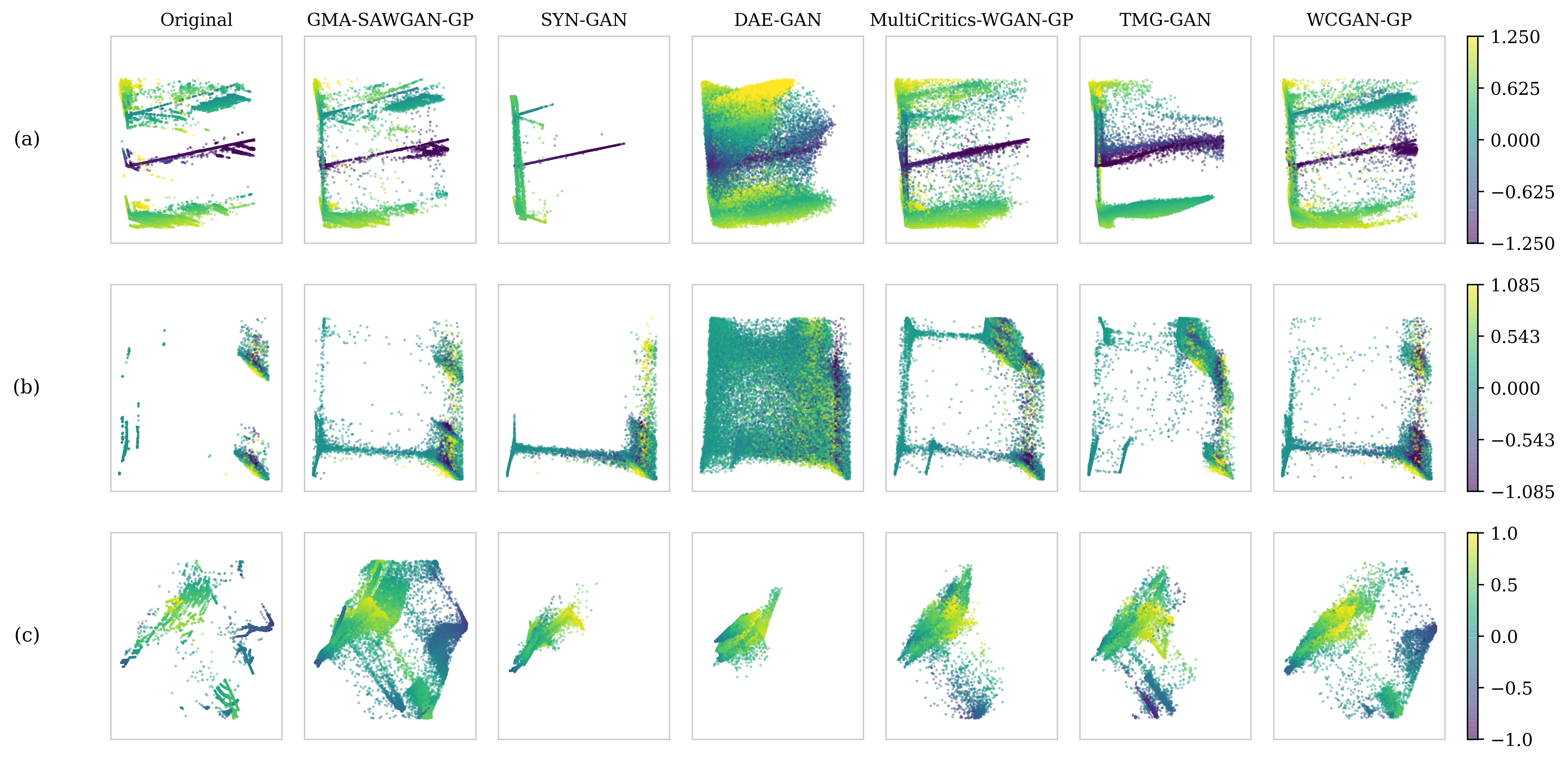}
  \caption{PCA projections of real and synthetic traffic across three datasets. Rows label (a)–(c) correspond to CICIDS2017, UNSW-NB15, and NSL-KDD, respectively. Columns show the original training data and synthetic samples generated by different GAN-based models. In each plot, the horizontal and vertical axes correspond to the first and second principal components of PCA. Colors indicate a standardized scalar, with scales shown on the right side.}
  \label{fig:all_datasets_gallery_wide}
\end{figure*}

The proposed GMA-SAWGAN-GP is compared with five state-of-the-art GAN-based methods designed to enhance IDS performance. The five GAN-based models include SYN-GAN \cite{rahman2024syn}, DAE-GAN \cite{9893038}, MultiCritics-WGAN-GP \cite{10852156}, TMG-GAN \cite{10312801}, and WCGAN-GP \cite{srivastava2023wcgan}. Fig. \ref{fig:all_datasets_gallery_wide} visualizes the geometric relationship between original network traffic and synthetic samples generated by these models using Principal Component Analysis (PCA) for each dataset by plotting samples along the first two principal components, enabling visual assessment of coverage. To ensure a fair comparison, all experiments employed the same preprocessed datasets, identical training procedures, and consistent evaluation settings with the number of epochs fixed at 5,000. 

For CICIDS2017, the original PCA projection reveals two compact high-density clusters along with several scattered branches. GMA-SAWGAN-GP reproduces both the compact clusters and peripheral branches, generating synthetic samples that interleave with the real data across the main manifolds. In contrast, SYN-GAN and WCGAN-GP tend to concentrate samples near a subset of clusters, while DAE-GAN fills low-density areas between them, altering the global geometry. MultiCritics-WGAN-GP and TMG-GAN capture the coarse location of the main clusters but generate fewer points in the more diffuse peripheral regions, indicating residual mode contraction for this dataset.

For UNSW-NB15, the original network data forms several elongated bands with clear low-density gaps. The synthetic samples from GMA-SAWGAN-GP closely follow these bands in both orientation and extent, populating dense cores and boundary regions, indicating strong coverage of the real data distributions. In contrast, SYN-GAN mainly generates points along a single dominant band, under-representing other regions, while DAE-GAN disperses samples into a relatively homogeneous rectangular area, smoothing over the thin structures presented in the original projection. MultiCritics-WGAN-GP, TMG-GAN, and WCGAN-GP approximate the overall outline of real clusters more effectively but still display uneven coverage, with oversampling in dense areas and visible sparsity along cluster boundaries.

For NSL-KDD, the original data occupies several curved, partially overlapping sheets. GMA-SAWGAN-GP and WCGAN-GP produce synthetic samples that align with these sheets and preserve their curvature, with color gradients similar to those of the real projection, indicating retention of intra-manifold variation. In contrast, SYN-GAN and DAE-GAN generate more compact clouds centered on only part of the original samples, while MultiCritics-WGAN-GP and TMG-GAN recover the general footprint but exhibit reduced spread and less detail along manifold edges.

Overall, the PCA visualizations suggest that GMA-SAWGAN-GP achieves the closest geometric alignment with real data across all three datasets, offering superior coverage of cluster boundaries and mitigating mode contraction compared to state-of-the-art GAN-based models.

Table \ref{tab:cmp_by_ids} compares the mean accuracies of the proposed generative model with those of five state-of-the-art approaches across 20 independent runs for five IDS models under both binary and multi-classification tests. The proposed method achieves the highest or ties for the highest mean accuracy in a substantial number of cases and generally ranks second or comparable in the remaining tests. Performance on NSL-KDD is overall weaker than on CICIDS2017 and UNSW-NB15, likely due to its coarse-grained categorical features, which become high-dimensional and sparse after encoding, limiting class separability and generative fidelity. Additionally, NSL-KDD exhibits extreme class imbalance; for example, the U2R class contains only 52 samples, introducing distributional shift during GAN training and data generation. Overall, the proposed GMA-SAWGAN-GP enhances IDS generalizability and captures diverse attack patterns more effectively than the competing methods.

\subsection{Ablation Study}

Table \ref{tab:ablation_experiment} presents the results of the ablation experiment. The baseline G-WGAN-GP uses Gumbel–Softmax regularization to handle discrete data. In GA-WGAN-GP, an AE reconstruction constraint is added to G-WGAN-GP, yielding notable improvements - particularly in multi-class accuracy for CNN-, CNN-BiLSTM-, and CNN-LSTM-based IDS models on UNSW-NB15 and CICIDS2017. GMA-WGAN-GP introduces a more powerful MLP and AE, along with feature alignment, to further stabilize these gains. The full GMA-SAWGAN-GP achieves the best or near-best results in most tests. In this configuration, the SA mechanism within the generator models cross-feature dependencies, while a lightweight entropy-based attention gate adaptively balances adversarial and reconstruction losses during training, improving stability and the quality of generated samples. These findings confirm that each element of GMA-SAWGAN-GP contributes meaningfully to enhancing IDS detection performance.

\begin{table*}[htbp]
\centering
\caption{Results of the Ablation Experiments}
\label{tab:ablation_experiment}

\begin{adjustbox}{center, max width=\textwidth}
\begin{threeparttable}
\scriptsize
    \begin{tabular}{l l C C C C C C}
      \toprule
      \multirow{2}{*}{IDS model} & \multirow{2}{*}{Ablation Model}
      & \multicolumn{2}{c}{UNSW\textendash NB15}
      & \multicolumn{2}{c}{NSL\textendash KDD}
      & \multicolumn{2}{c}{CICIDS2017} \\
      \cmidrule(lr){3-4}\cmidrule(lr){5-6}\cmidrule(lr){7-8}
      & & B.Acc (\%) & M.Acc (\%) & B.Acc (\%) & M.Acc (\%) & B.Acc (\%) & M.Acc (\%) \\
      \midrule
      \multirow{4}{*}{CNN}
      & G-WGAN-GP         & 83.0$\pm$2.3 & 86.5$\pm$0.8 & 78.9$\pm$1.3 & 77.4$\pm$0.8 & 93.3$\pm$1.4 & 97.3$\pm$0.2 \\
      & GA-WGAN-GP & 83.8$\pm$1.2 & 87.8$\pm$0.7 & 79.0$\pm$1.3 & 76.8$\pm$0.3 & 97.3$\pm$0.9 & 97.3$\pm$0.2 \\
      & GMA-WGAN-GP     & 85.9$\pm$3.7 & 87.3$\pm$0.5 & 82.1$\pm$2.5 & 76.8$\pm$0.3 & 97.3$\pm$0.6 & 97.3$\pm$0.2\\
      & GMA-SAWGAN-GP &  89.4$\pm$2.3 &89.4$\pm$0.3 & 81.8$\pm$2.3 & 78.8$\pm$0.7 & 97.3$\pm$0.6 & 97.4$\pm$0.4 \\
      \midrule
      \multirow{4}{*}{DNN}
      & G-WGAN-GP         & 84.6$\pm$1.3 & 83.7$\pm$1.1  & 78.0$\pm$0.9 & 74.4$\pm$0.5 & 91.2$\pm$0.6 & 96.5$\pm$0.4 \\
      & GA-WGAN-GP     & 84.8$\pm$1.3 & 85.7$\pm$1.7 & 79.0$\pm$0.6 & 74.4$\pm$0.6 & 97.2$\pm$0.2 & 96.8$\pm$0.2 \\
      & GMA-WGAN-GP     & 87.6$\pm$0.8 & 84.7$\pm$1.2 & 78.1$\pm$0.8 & 74.4$\pm$0.6 & 97.1$\pm$0.3 & 96.6$\pm$0.3 \\
      & GMA-SAWGAN-GP & 88.3$\pm$0.8 & 85.7$\pm$1.7 & 79.0$\pm$0.6 & 74.2$\pm$0.2 & 97.2$\pm$0.2 & 96.7$\pm$0.3 \\
      \midrule
      \multirow{4}{*}{LSTM}
      & G-WGAN-GP         & 87.6$\pm$1.2 & 85.5$\pm$1.2 & 80.6$\pm$0.7 & 76.0$\pm$0.8 & 92.9$\pm$0.8 & 97.3$\pm$0.1 \\
      & GA-WGAN-GP & 87.6$\pm$1.8 & 86.1$\pm$0.8 & 79.0$\pm$1.1 & 75.7$\pm$0.5 & 97.0$\pm$0.2 & 97.2$\pm$0.1 \\
      & GMA-WGAN-GP& 87.6$\pm$0.8 & 84.7$\pm$1.2 & 79.0$\pm$1.1 & 75.7$\pm$0.5 & 97.2$\pm$0.1 & 97.3$\pm$0.0 \\
      & GMA-SAWGAN-GP & 88.3$\pm$0.9 & 85.1$\pm$0.9 & 84.6$\pm$1.2 & 79.6$\pm$0.8 & 97.1$\pm$0.2 & 97.3$\pm$0.2 \\
      \midrule
      \multirow{4}{*}{CNN-BiLSTM}
      & G-WGAN-GP         & 85.6$\pm$0.9 & 83.4$\pm$1.5 & 82.4$\pm$0.8 & 75.1$\pm$1.5 & 98.2$\pm$0.1 & 98.2$\pm$0.0 \\
      & GA-WGAN-GP     & 86.3$\pm$1.5 & 83.8$\pm$2.1 & 78.0$\pm$0.6 & 75.1$\pm$1.5 & 94.1$\pm$0.8 & 98.0$\pm$0.2 \\
      & GMA-WGAN-GP    & 86.2$\pm$0.8 & 89.5$\pm$0.5 & 78.0$\pm$0.6 & 76.2$\pm$1.2 & 98.1$\pm$0.2 & 98.0$\pm$0.1 \\
      & GMA-SAWGAN-GP & 90.0$\pm$0.6 & 87.6$\pm$0.8 & 84.4$\pm$1.9 & 79.8$\pm$1.2 & 98.1$\pm$0.1 & 98.1$\pm$0.1  \\
      \midrule
      \multirow{4}{*}{CNN-LSTM}
     & G-WGAN-GP         & 87.5$\pm$1.0 & 84.5$\pm$0.9 & 81.9$\pm$0.5 & 76.9$\pm$0.7 & 98.5$\pm$0.3 & 98.3$\pm$0.0 \\
      & GA-WGAN-GP  & 87.9$\pm$0.7 & 85.8$\pm$0.7 & 81.8$\pm$0.4 & 77.3$\pm$0.5 & 94.2$\pm$0.5 & 98.3$\pm$0.0 \\
      & GMA-WGAN-GP    & 89.5$\pm$0.5 & 86.4$\pm$1.7 & 76.7$\pm$0.2 & 77.0$\pm$0.6 & 98.5$\pm$0.2 & 98.3$\pm$0.0 \\
      & GMA-SAWGAN-GP & 89.7$\pm$1.1 & 86.4$\pm$0.4 & 82.9$\pm$0.6 & 81.4$\pm$0.7 & 99.2$\pm$0.2 & 98.4$\pm$0.0  \\
      \bottomrule
\end{tabular}

\begin{tablenotes}
\scriptsize
\item \textit{Note:} Results are Binary classification Accuracy (B.Acc) and Multi-classification Accuracy (M.Acc).
\end{tablenotes}
\end{threeparttable}
\end{adjustbox}

\end{table*}

\subsection{Study Limitations}

This study has the following limitations. First, all experiments were conducted on NSL-KDD, UNSW-NB15, and CICIDS2017, which, despite their widespread use, contain historical traffic and idealized features and cannot fully replicate real-world traffic conditions or attack-defence dynamics \cite{hesford2024expectations}. Consequently, IDS models trained on legacy traffic may experience significant performance degradation when deployed in contemporary environments due to distributional shifts. Second, the evaluation includes five IDS models but excludes emerging approaches such as Transformer-based models and graph- or flow-based detectors; therefore, the conclusions may not generalize to all IDS paradigms. Third, although AUROC and TPR@5\%FPR are reported, the study does not comprehensively analyze cost-sensitive metrics, precision, and recall behavior under extreme imbalance, or the robustness of post-training calibration, which limits interpretability for unknown attack detection. Finally, in the comparative experiments, four GAN-based data generative models from prior literature experienced mode collapse during training and were excluded from analysis. While this filtering was necessary to ensure fairness, it may introduce bias by favoring methods that train stably under our experimental setup.

\section{Conclusion}

This paper addresses a common shortage of GAN-based augmentation in IDS research: many existing methods either apply one-hot encoding to discrete features or discard them, resulting in excessive sparsity and exacerbating vanishing gradients for mixed-type traffic. To overcome this, we propose GMA-SAWGAN-GP, a novel generative augmentation framework for mixed-type network data that integrates Gumbel–Softmax regularization for discrete feature modeling, an autoencoder for manifold alignment, and an MLP with self-attention for feature-level dependency modeling, combined with an adaptive loss gate for stable optimization. 

Extensive experiments on three widely used datasets - NSL-KDD, UNSW-NB15, and CICIDS2017 - across five IDS models demonstrate that GMA-SAWGAN-GP consistently enhances detection performance on known attacks, outperforming state-of-the-art intrusion detection methods in most binary and multi-classification tests. Furthermore, LOAO experiments confirm that the proposed approach enhances IDS generalization to unknown attacks, achieving higher AUROC and TPR@5\%FPR compared to baselines. Ablation studies further validate that each component of the framework contributes meaningfully to improving IDS detection performance. In summary, these results demonstrate that the proposed model enhances IDS detection of known attacks and generalizability to unknown attacks by generating high-quality synthetic traffic that expands coverage and stabilizes training.

In the future, we plan to explore alternative representations of network traffic to improve intrusion detection on small or sparse datasets. Such a transformation could leverage structural and contextual features, further enhancing IDS detection capability and generalization overall.

\section*{Acknowledgments}
The authors would like to acknowledge the Engineering and Physical Sciences Research Council (EPSRC), UK, for the support of this work (EP/W00366X/1).

\bibliographystyle{IEEEtran} 
\bibliography{references}

\vfill

\end{document}